\documentclass[journal]{IEEEtran}

\ifCLASSINFOpdf
\else
\fi
\usepackage{xcolor}
\usepackage{amsmath}
\usepackage{amsfonts}
\usepackage{amssymb}
\usepackage{amsthm}
\usepackage{graphicx}
\usepackage{caption}
\usepackage[font=footnotesize]{caption}
\usepackage[caption=false]{subfig}
\usepackage{lipsum}
\usepackage{cite}
\usepackage{amsthm}
\usepackage{morefloats}
\usepackage{algorithm} 
\usepackage{algpseudocode} 
\usepackage{algorithmicx}
\usepackage{hyperref} 
\usepackage{cancel}

\newcommand{\I}{{\bf I}}

\newcommand{\Eta}{{\boldsymbol{\eta}}}
\newcommand{\mtDr}{\tilde{{\mathcal{D}}}^{r\times r}}
\newcommand{\mtDk}{{\mathcal{D}}^{k\times k}}

\newcommand{\bthe}{{\boldsymbol{\theta}}}

\newcommand{\diag}{{\rm{diag}}}
\newcommand{\deter}{{\rm{det}}}

\makeatletter
\newcommand*{\rom}[1]{\expandafter\@slowromancap\romannumeral #1@}
\makeatother

\begin{document}

\title{Joint RIS Phase Profile Design and Power Allocation for Parameter Estimation in Presence of Eavesdropping}
\author{Erfan Mehdipour Abadi, Ayda Nodel Hokmabadi, and Sinan Gezici, \textit{Senior Member, IEEE}\thanks{E. M. Abadi is with Wayne State University, Detroit, Michigan 48202, USA. A. N. Hokmabadi and S. Gezici are with the Department of Electrical and Electronics Engineering, Bilkent University, Bilkent, Ankara 06800, Turkey, Tel: +90 (312) 290-3139, Fax: +90 (312) 266-4192, (e-mails: mehdipour.erfan96@gmail.com, \{nodel,gezici\}@ee.bilkent.edu.tr).}\vspace{-0.6cm}}

\maketitle

\begin{abstract}
We consider secure transmission of a deterministic complex-valued parameter vector from a transmitter to an intended receiver in the presence of an eavesdropper in a reconfigurable intelligent surface (RIS)-integrated environment. We aim to jointly optimize the RIS phase profile and the power allocation matrix at the transmitter to enhance the estimation accuracy at the intended receiver while limiting that at the eavesdropper. We utilize the trace of the Fisher information matrix (FIM), equivalently, the average Fisher information, as the estimation accuracy metric, and obtain its closed form expression for the intended receiver and the eavesdropper. Accordingly, the joint RIS phase profile and power allocation problem is formulated, and it is solved via alternating optimization. When the power allocation matrix is fixed during alternating optimization, the optimal RIS phase profile design problem is formulated as a non-convex problem and it is solved via semidefinite relaxation and rank reduction. When the RIS phase profile is fixed, a linear programming formulation is obtained for optimal power allocation. Via simulations, the effects of RIS phase design and power allocation are illustrated individually and jointly. Moreover, extensions are provided by considering the presence of line of sight paths in the environment and the availability of RIS elements with adjustable magnitudes.


% \fi
% For peer-review papers, this IEEEtran command inserts a page break and
%\IEEEpeerreviewmaketitle
\end{abstract}

\vspace{-0.3cm}

\section{Introduction}\label{sec:intro}

\subsection{Literature Review}

Security, cost-effectiveness, and energy efficiency are fundamental needs of beyond fifth-generation (B5G) and sixth-generation (6G) systems \cite{RIS_6G1,RIS_6G2,RIS_6G3}. Due to the broadcasting nature of wireless channels and security vulnerability in wireless networks, % such as IoT, WSN, etc.,
eavesdropping is one of the significant concerns related to the security of B5G and 6G systems \cite{RIS_6G4,IoT,WSN}. Key-based cryptographic approaches are extensively studied to provide secrecy in the literature. Meanwhile, heterogeneous and dynamic networks cause a variety of challenges and cost-inefficiency in terms of key generation and distribution in these methods \cite{crypto1,crypto2}. 
For cost, power, latency, and bandwidth limitations of massive device networks, physical layer security (PLS) has attracted significant attention as an alternative to traditional approaches due to its information-theoretic security base and cryptography-free characteristics \cite{PLS1,PLS2}. The key idea behind PLS, first proposed by Wyner \cite{Wyner}, is to securely transmit parameters to intended (legitimate) receivers in presence of eavesdroppers by exploiting their wireless channel characteristics \cite{PLS3}.

In order to improve security with low power consumption and in a cost-efficient way, reconfigurable intelligent surfaces (RISs) have recently attracted significant research attention. RIS is a planar surface with a programmable structure equipped with low-cost and passive meta-surface components. An RIS can alter the propagation direction of electromagnetic waves by appropriately shifting the phases of signals without any additional equipment such as RF processing \cite{RIS1,RIS2}, whereby the security can be increased by manipulating the channel characteristics of the intended receiver and the eavesdropper \cite{RIS_SR1,RIS_SR2,RIS_SR3,RIS_SR4,RIS_SR8,RIS_SR7,RIS_SR12,RIS_SR13,RIS_SR14,RIS_SR10,RIS_SR11,RIS_SR5,RIS_SR6,RIS_SR9}.
For PLS purposes, different types of metrics, such as information, detection, or estimation theoretic metrics, are used in the literature to quantify the amount of secrecy. 

Information theoretic metrics commonly consist of mutual information, secrecy rate (SR), secrecy outage probability (SOP), and secrecy capacity (SC). \cite{Wyner} shows that a non-zero SR is achievable at the intended receiver providing zero information to the eavesdropper if the channel of the eavesdropper is a degraded version of that of the intended receiver. In addition, \cite{SR1} and \cite{SR2} utilize artificial noise (AN) and beamforming in order to maximize the secrecy communication rate. In \cite{RIS_SR1,RIS_SR2,RIS_SR3,RIS_SR4,RIS_SR8,RIS_SR7}, integration of RIS into a multi-input single-output (MISO) system is studied for maximizing the achievable SR, SR, minimum SR, and weighted sum SR with the combination of linear precoding, beamforming, and power allocation by considering delay-limitations for quality of service (QoS). \cite{RIS_SR12} and \cite{RIS_SR13} aim to minimize the transmitted power and maximize the harvested power of the receiver subject to the SR constraint, while \cite{RIS_SR14} focuses on minimizing the transmitted power subject to SR and SOP constraints. Moreover, in \cite{RIS_SR10}, achievable SR is maximized with AN, precoding, and RIS phase shifts in a multiple-input multiple-output (MIMO) system, and \cite{RIS_SR11} performs joint optimization of beamforming and RIS phase shifts in order to maximize the SR in a MIMO system. Also, \cite{RIS_SR5} aims to maximize the achievable ergodic rate under the QoS constraint by joint optimization of AN and RIS phase shifts in a MIMO system. Meanwhile, the ergodic secrecy rate in a single-input single-output (SISO) system is maximized in \cite{RIS_SR6} via optimization of RIS phase shifts. Furthermore, the average worst-case SR of a simultaneous wireless information and power transfer (SWIPT) system is investigated in \cite{RIS_SR9} for joint optimization of beamforming and RIS phase shifts. SOP is adopted as a secrecy metric in various settings such as 
\cite{SOP1,SOP2,SOP3}, and the studies in \cite{SOP_RIS2,SOP_RIS3,SOP_RIS4,SOP_RIS5} consider integration of RIS into SISO systems and derive analytical expressions for SOP, average SR, and non-zero SR. Besides, \cite{SOP_RIS1} formulates the SOP minimization problem via joint optimization of beamforming and RIS phase shifts in multiple-input multiple-output multiple-antenna-eavesdropper (MIMOME) channels. SC is another secrecy metric that is utilized in the literature. For example, \cite{SC_RIS3} considers the SC in a RIS-integrated SISO system in a measurement-based manner. Also, \cite{SC_RIS2} formulates the average SC in a RIS-integrated SISO system and \cite{SC_RIS1} adjusts the SC for a high-speed railway system for joint beamforming and RIS phase shift design. 

For secure information transmission, estimation theoretic metrics, such as Fisher information, mean-squared error (MSE), and Cram\'er-Rao lower bound (CRLB), are also employed in various work in the literature. CRLB provides a lower bound on MSEs of unbiased estimators and is derived from the inverse of the Fisher information matrix (FIM) \cite{poor_book}. In \cite{CRLB1}, a secure inference problem is introduced for an Internet of Things (IoT) network with sufficient data for deterministic parameter estimation under spoofing and man-in-the-middle attack (MiMA) based on the CRLB metric. In \cite{CRLB2, CRLB3, CRLB4, CRLB5, TsP2022}, secure estimation of a random parameter is investigated in the presence of an eavesdropper in various scenarios and setups. In particular, in \cite{CRLB2} and \cite{CRLB4}, minimization of the expectation of the conditional Cram\'er-Rao lower bound (ECRB) at the intended receiver is performed by optimizing the deterministic encoding at the transmitter for random scalar and vector parameters under the condition that the linear minimum mean-squared error (LMMSE) of the eavesdropper is above a specific secrecy bound. Moreover, \cite{CRLB3} proposes the worst-case CRLB of the parameter as a performance metric for achieving robust parameter encoding. In \cite{CRLB5}, optimal power allocation (OPA) for secure estimation of vector parameters is proposed under a total power constraint while keeping the MSE of the eavesdropper above a certain value by assuming the use of the maximum likelihood (ML) estimator at the eavesdropper. In the aforementioned studies, it is assumed that the eavesdropper is not aware of encoding at the transmitter. On the other hand, in \cite{TsP2022}, OPA and optimal linear encoding (OLE) are proposed by assuming that the eavesdropper is aware of the encoding at the transmitter (i.e., a smart eavesdropper), which presents a worst-case scenario for the estimation performance of the intended receiver. By considering the CRLB and the determinant of the FIM as the estimation accuracy metrics, OPA and OLE are performed under power and secrecy constraints in \cite{TsP2022}. In another line of work, CRLB is used in various RIS-integrated setups for improvement of system performance in the absence of eavesdropping. In \cite{CRLB_RIS1}, the goal is to minimize the multi-user interference (MUI) in a MISO system, while RIS elements have constant-modulus and discrete phase values under the CRLB constraint for direction of arrival (DOA) estimation. In addition, \cite{CRLB_RIS2} investigates joint optimization of beamforming and RIS phase profile for minimizing the CRLB of point and extended targets. Investigation of FIM and CRLB for RIS-integrated systems requires analysis for complex parameters, and the derivation of FIM and CRLB for complex parameters can be found in \cite{complexFIM2,complexFIM3}. 
%To the best of our knowledge, the secrecy of RIS-integrated systems has not been studied based on CRLB and FIM metrics.    

Estimation theoretic metrics for physical layer security are not limited to the CRLB. Other metrics, such as total MMSE, average MSE, and Fisher information, are also used to quantify the secrecy of various systems \cite{nonCRLB1,nonCRLB2,nonCRLB3}. Specifically, in \cite{nonCRLB2}, a Bayesian framework with a Gaussian prior distribution for both intended receivers and eavesdroppers is considered, where the total MMSE of the intended receivers is minimized under a constraint on the MMSE at the eavesdroppers using joint AN and linear encoding. Additionally, for distributed parameter estimation, optimal transmit power allocation policies are derived in \cite{nonCRLB3} when the employed metrics are the average MSEs of the parameter of interest at the eavesdroppers. Furthermore, \cite{nonCRLB1} adopts a non-Bayesian parameter estimation framework for ensuring the privacy of smart meter data in a smart grid network based on the Fisher information metric, namely, the Fisher information for a single variable and the trace of the FIM for the multi-variable parameter estimation problem. Moreover, in \cite{nonCRLB4, trFIM_criterion}, the trace and the log-determinant of the Bayesian FIM are introduced as other useful metrics for estimation problems without considering any eavesdroppers in the system. 

\subsection{Contributions}

Secure parameter transmission problems have been investigated in various studies in the literature, e.g., \cite{CRLB2,CRLB3,CRLB4,CRLB5,TsP2022,nonCRLB2}, by utilizing various estimation theoretic metrics, such as the CRLB and MSE, and considering both random and deterministic parameters. In these studies, optimal parameter encoding and optimal power allocation problems are formulated for enhancing the estimation accuracy at an intended receiver while imposing a constraint on estimation accuracy of eavesdropping. 
%In spite of the investigations conducted in \cite{CRLB2,CRLB3,CRLB4,CRLB5}, which explore the use of estimation theoretic metrics within a Bayesian framework to address optimal parameter encoding issues for secure transmission of random parameters in the presence of an unaware eavesdropper, \cite{TsP2022} presents optimal power allocation and optimal linear encoding techniques for secure parameter transmission when dealing with a \textit{smart} eavesdropper who is knowledgeable about the encoding schemes. 
A recent advancement in communication systems is the introduction of RIS, which can play a vital role in system performance improvement through proper control of its elements. Although secure parameter transmission problems have been investigated in the aforementioned studies from an estimation theoretic perspective, there exists no work in the literature that considers the presence of an RIS in a secure parameter transmission system and performs joint optimization of RIS phase profile and power allocation based on estimation theoretic metrics.
%Considering that RIS can induce predetermined changes in the parameter phase, it becomes essential to study complex-valued parameter estimation for systems integrated with RIS, which requires further investigation from an estimation theoretic point of view.

In this manuscript, we propose a secure parameter transmission problem in the presence of an RIS in order to enhance estimation accuracy of an intended receiver while keeping the estimation accuracy of an eavesdropper below a certain limit. To this aim, we formulate a problem that involves the joint optimization of power allocation and RIS phase profile according to the average Fisher information metric. Namely, by considering a deterministic and complex-valued unknown parameter vector, the objective is to maximize the trace of the FIM at the intended receiver under the constraints that the trace of the FIM is below a certain limit for the eavesdropper, the reflection coefficients of the RIS elements have unit amplitudes, and the total transmit power is limited by a certain value. 
%The objective is to achieve secure transmission of a \textit{deterministic} complex-valued parameter vector in the presence of a \textit{smart} eavesdropper who has knowledge of the power allocation function and RIS phase profile determined by the transmitter. To address this, we utilize a metric based on the FIM to quantify the estimation performance of both the legitimate receiver and the eavesdropper, considering the unit modulus constraint of RIS elements and the total power constraint at the transmitter and constraint of estimation performance at the eavesdropper. Specifically, we focus on the average Fisher information, which is the trace of the FIM, as the performance metric. 
By deriving a closed form expression for the trace of the FIM, we formulate the joint power allocation and RIS phase profile design problem, and propose an alternating optimization approach to solve it. During alternating optimization, when power allocation is fixed, the optimal RIS phase profile design problem is formulated as a non-convex problem, and it has been solved via semidefinite relaxation (SDR)  and rank reduction. When the RIS phase profile is fixed, a linear programming formulation is obtained for optimal power allocation. Although it is initially assumed that the transmitter is in non-line-of-sight (NLoS) with both the intended receiver and the eavesdropper, extensions are also provided by considering the presence of line-of-sight paths in the environment. Moreover, the problem formulation is also extended to the scenario when the RIS elements have reflection coefficients with adjustable magnitudes. The key contributions and novelty of this manuscript can be outlined as follows:
\begin{itemize}
    \item For the first time in the literature, we propose a joint RIS phase profile design and power allocation problem for secure transmission of a complex-valued parameter vector in an RIS-integrated environment to maximize the average Fisher information at the intended receiver while ensuring that the average Fisher information at the eavesdropper remains below a predefined threshold. These objectives are accomplished under the total transmit power constraint and the requirement of unit-modulus reflection coefficients for the RIS elements.
    \item FIMs are explicitly derived for the intended receiver and the eavesdropper regarding the estimation of a complex-valued parameter vector in an RIS-integrated environment.
    \item An alternating optimization solution is proposed for obtaining the (sub)optimal power allocation and RIS phase profile. During alternating optimization, for fixed power allocation, the optimal RIS phase profile design problem is formulated as a non-convex problem, which is solved via SDR and rank reduction. On the other hand, for a fixed RIS phase profile, a linear programming formulation is obtained for optimal power allocation.  
    \item It is shown that the results obtained for the NLoS scenario can also be extended to cases in the presence of LoS paths in the environment via homogenizing the optimization problem developed for the NLoS scenario. 
    \item It is shown that the unit modulus constraint imposed on the RIS elements can lead to infeasibility problems in some cases, particularly when the secrecy constraint is set to low values. However, when the RIS elements have reflection coefficients with adjustable magnitudes, this infeasibility issue is resolved and the problem formulation is extended to cover this scenario, as well.
    %To overcome these issues, two potential solutions are proposed. Firstly, the power allocation approach in the alternating optimization problem can be modified, but this is only feasible if an encoding scheme is accessible at the transmitter, alternatively, by increasing the degree of freedom and gaining control over the magnitude of RIS element reflections.
\end{itemize}

%\vspace{-0.2cm}

\subsection{Organization}

The remainder of the manuscript is structured as follows. Section~\ref{sec:sys} provides a description of the system model and introduces the problem formulation. In Section~\ref{optimization}, the proposed approaches for RIS phase profile design and power allocation are developed by considering the average Fisher information criterion. Section~\ref{sec:Nume} presents various numerical examples for investigating the performance of the proposed approaches. Section \ref{sec:extensions} extends the theoretical results to account for the presence of dominant LoS components and RIS elements with adjustable magnitudes. Finally, concluding remarks are made in Section \ref{sec:conclusion}.

\section{System Model and Problem Formulation}\label{sec:sys}

%\textcolor{red}{ system structure, photo, matrices names, and sizes ... }

\begin{figure}[ht]
    \includegraphics[width=0.9\linewidth]{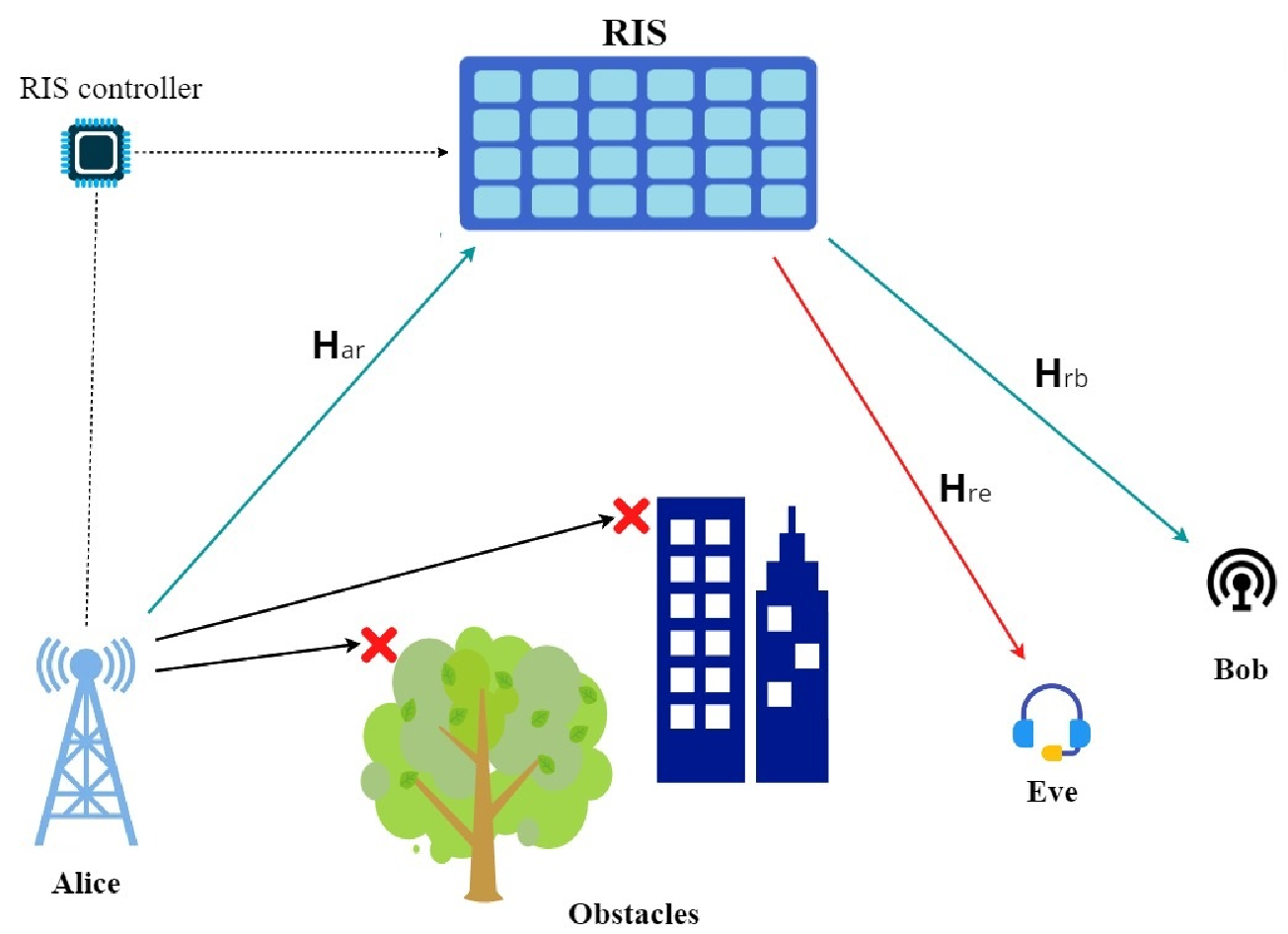}
    \centering
    \captionsetup{width=\linewidth}
    \caption{RIS-assisted system in the presence of an eavesdropper.}
    \label{RIS-assisted}
\end{figure}

A $k$-dimensional complex parameter vector denoted by ${\bthe} = [\theta_{1}, \theta_{2}, \dots, \theta_{k}]^{T} \in \mathbb{C}^{k}$ is to be sent from a transmitter, Alice, to an intended receiver, Bob, in the presence of an eavesdropper, Eve, when the LoS paths are blocked by obstacles, as depicted in Fig~\ref{RIS-assisted}. By assuming no prior statistical information about $\bthe$ at the intended receiver and the eavesdropper, $\bthe$ is modeled as a deterministic unknown parameter vector for Bob and Eve. The aim is to improve the 
%average Fisher information in the received signal 
parameter estimation performance at Bob while keeping that below a particular limit at Eve by designing a proper RIS phase profile in the smart radio environment and performing optimal power allocation at the transmitter (Alice). It is assumed that the transmitter has a reliable connection to the RIS for adjusting the phases of the RIS elements, as shown in Fig.~\ref{RIS-assisted}. %\textcolor{green}{(Any name for the arrow?)}
    
Measurements are obtained at Bob and Eve via the following linear models \cite{RIS_SR2, LinearModel2}:
\begin{align} \label{y_bob}
    {\bf{y}}_b&=\left( {\bf{H}}_{rb}{\bf{\Omega}}{\bf{H}}_{ar} \right){\bf{P}}{\bthe}+\Eta_b\\ \label{y_eve}
    {\bf{y}}_e&=\left( {\bf{H}}_{re}{\bf{\Omega}}{\bf{H}}_{ar} \right){\bf{P}}{\bthe}+\Eta_e
\end{align}
where ${\bf{y}}_b \in \mathbb{C}^{n_b}$ and  ${\bf{y}}_e \in \mathbb{C}^{n_e}$ denote the measurements at Bob and Eve, respectively. Also, 
${\bf{H}}_{ar} \in \mathbb{C}^{r\times k}$, ${\bf{H}}_{rb}\in \mathbb{C}^{n_b\times r}$, and ${\bf{H}}_{re}\in \mathbb{C}^{n_e\times r}$ are channel coefficient matrices for Alice-RIS, RIS-Bob, and RIS-Eve channels, respectively, and $k$, $n_b$, $n_e$, and $r$ are the size of the parameter vector, the number of antenna elements at Bob, the number of antenna elements at Eve, and the number of RIS elements, respectively. Also, $\Eta_b \in \mathbb{C}^{n_b}$ and $\Eta_e \in \mathbb{C}^{n_e}$ are additive circularly-symmetric Gaussian noise vectors at Bob and Eve, with distributions 
${\Eta}_b \sim \mathcal{CN}(\boldsymbol{0},\boldsymbol{\Sigma}_{b})$ and 
${\Eta}_e \sim \mathcal{CN}(\boldsymbol{0},\boldsymbol{\Sigma}_{e})$, respectively, where
$\boldsymbol{\Sigma}_{b}$ and
$\boldsymbol{\Sigma}_{e}$ are the covariance matrices, which are assumed to be positive definite. In addition, ${\bf{P}} = \diag \{ \sqrt{p_1},\sqrt{p_2}, \dots,\sqrt{p_k}  \}$ and ${\bf{\Omega}}= \diag \{ \omega_1,\omega_2, \dots, \omega_r \}$ are, respectively, the power allocation matrix and the RIS phase profile to be optimized, where $p_j$'s are non-negative real numbers for $j=1,2,\dots,r$ and $\omega_i = e^{j \phi_i}$ 
%where $\phi_i \in [0,2 \pi)$ 
for $i=1,2,\dots,r$ with $\phi_i$ denoting the phase of the $i$th RIS element.

%\textcolor{red}{     motivations and supporting ideas(What is trace of FIM for and how it helps us)}

In this study, the aim is to design the RIS phase profile, ${\bf{\Omega}}$, and the power allocation matrix, ${\bf{P}}$, for enhancing the estimation performance at the intended receiver, Bob, while keeping the estimation performance at the eavesdropper, Eve, below a particular limit. For quantifying estimation performance, Fisher information based optimality criteria can be preferred as they do not depend on specific estimator structures and lead to compact expressions \cite{Emery_1998,Dogancay1,trFIM_criterion,kalaba,tzoreff}. The most commonly used Fisher information based optimality criteria are the trace of the inverse FIM, i.e., the Cram\'{e}r-Rao lower bound (CRLB), the log-determinant of the FIM, the maximum eigenvalue of the CRLB, the largest diagonal entry of the CRLB, the trace of the FIM, and the minimum diagonal element of the FIM \cite{Emery_1998,Dogancay1,trFIM_criterion}.\footnote{The CRLB, the log-determinant of the FIM, the maximum eigenvalue of the CRLB, and the largest diagonal entry of the CRLB are also referred as A-optimality, D-optimality, E-optimality, and G-optimality criteria, respectively \cite{Emery_1998,Dogancay1,trFIM_criterion}.} Among these optimality criteria, we employ the trace of the FIM (i.e., the average Fisher information criterion) in this study due to two main reasons: ($i$) Since the informativeness of the measurement for estimating the $i$th element of the parameter vector corresponds to the $i$th diagonal entry of the FIM, the trace of the FIM indicates the overall usefulness of the measurement for vector parameter estimation \cite{trFIM_criterion}. ($ii$) For the considered problem, the trace of the FIM can be derived in closed form, which facilitates theoretical analyses and provides intuition.

%\textcolor{red}{optimization criterion optimization problem tr\{FIM\}}

Let ${\I}\left({{\bf{y}}_b};\bthe \right)$ and ${\I}\left({{\bf{y}}_e};\bthe \right)$, denote, respectively, the FIMs corresponding to measurements ${\bf{y}}_b$ and ${\bf{y}}_e$, with respect to the complex parameter vector $\bthe$ (which will be derived in Section~\ref{optimization}). Based on these FIMs and by considering the trace of FIM as the performance metric, the following problem formulation is proposed. 
\begin{subequations}\label{eq:optProblem}
\begin{align}\label{eq:optProblema}
    \max_{{\bf{P}}\in\mtDk,\,{\bf{\Omega}}\in\mtDr}\quad &{\rm{tr}}\{{\bf{I}}({\bf{y}}_b;{\bthe})\}\\\label{eq:optProblemb}
    {\rm{s.t.}}\hspace{0.9cm}\quad & {\rm{tr}}\{{\bf{I}}({\bf{y}}_e;{\bthe})\} \leq \Delta  \\\label{eq:optProblemc}
    \quad &{\rm{tr}}\{{\bf{P}}{\bf{P}}^T\}\leq P_\Sigma\\\label{eq:optProblemd}
    \quad & {\bf{P}}{\bf{P}}^T\succeq {\bf{0}}\\\label{eq:optProbleme}
    \quad & \big{|}[{\bf{\Omega}}]_{\ell,\ell}\big{|}=1, ~~\ell=1,\ldots,r
\end{align}
\end{subequations}
where $\mtDk$ denotes the set of $k\times k$ real diagonal matrices, $\mtDr$ represents the set of $r\times r$ complex diagonal matrices, (\ref{eq:optProblemb}) corresponds to the secrecy constraint with $0 \leq \Delta  < \infty$ denoting the secrecy limit, (\ref{eq:optProblemd}) and (\ref{eq:optProblemc}) impose, respectively, non-negativity of power levels and the total power constraint at Alice with $0 < P_\Sigma < \infty$ representing the power limit, and (\ref{eq:optProbleme}) states the unit amplitude nature of the RIS phase profile.

%\textcolor{red}{Alternating optimization, RIS phase profile, Power allocation}

As \eqref{eq:optProblem} is very challenging to solve directly, we employ an alternative optimization approach \cite{RIS_SR2, SOP_RIS1, RIS_SR12, RIS_SR1,AO_notes}, and divide the problem into two separate problems. First, the optimal RIS phase profile is designed by considering a given power allocation matrix based on the following problem:
\begin{subequations}\label{eq:RISomegaProblem}
\begin{align}\label{eq:RISomegaProblema}
    \max_{{\bf{\Omega}}\in\mtDr}\quad &{\rm{tr}}\{{\bf{I}}({\bf{y}}_b;{\bthe})\}\\\label{eq:RISomegaProblemb}
    {\rm{s.t.}}\quad\,\,\,\, & {\rm{tr}}\{{\bf{I}}({\bf{y}}_e;{\bthe})\} \leq \Delta  \\\label{eq:RISomegaProblemc}
    \quad & \big{|}[{\bf{\Omega}}]_{\ell,\ell}\big{|}=1, ~~\ell=1,\ldots,r
\end{align}
\end{subequations}
Then, the optimal power allocation matrix is derived for a given RIS phase profile via the following formulation:
\begin{subequations}\label{eq:PAoptProblem}
\begin{align}\label{eq:PAoptProblema}
    \max_{{\bf{P}}\in\mtDk}\quad &{\rm{tr}}\{{\bf{I}}({\bf{y}}_b;{\bthe})\}\\\label{eq:PAoptProblemb}
    {\rm{s.t.}}\quad\,\,\,\, & {\rm{tr}}\{{\bf{I}}({\bf{y}}_e;{\bthe})\} \leq \Delta  \\\label{eq:PAoptProblemc}
    \quad &{\rm{tr}}\{{\bf{P}}{\bf{P}}^T\}\leq P_\Sigma\\\label{eq:PAoptProblemd}
    \quad & {\bf{P}}{{\bf{P}}^T}\succeq {\bf{0}}
\end{align}
\end{subequations}
We can obtain the solution of (\ref{eq:optProblem}) via an alternating optimization approach by solving the problems in (\ref{eq:RISomegaProblem}) and (\ref{eq:PAoptProblem}) iteratively. Detailed investigations are provided in the following sections.
%while keeping the other parameters fixed.

%\textcolor{red}{  introduces scenarios and extensions(LoS, Discrete). ....}

%------------------------------------------------------------------

\section{RIS Phase profile Design and Power Allocation based on Average Fisher Information Criterion}\label{optimization}

In this section, we first derive the FIM and calculate the trace of the FIM explicitly. Then, we obtain the sub-optimal RIS phase profile and optimal power allocation matrix in different subsections. Finally, we focus on the joint alternating optimization problem in the last part.

According to (\ref{y_bob}) and (\ref{y_eve}), the conditional distributions of the measurements at Bob and Eve can be written as ${{\bf{y}}_b}\,|\,\bthe \sim \mathcal{CN}(({\bf{H}}_{rb}{\bf{\Omega}}{\bf{H}}_{ar} ){\bf{P}}{\bthe},{{\bf{\Sigma}}_b})$ and ${{\bf{y}}_e}\,|\,\bthe \sim \mathcal{CN}(({\bf{H}}_{re}{\bf{\Omega}}{\bf{H}}_{ar} ){\bf{P}}{\bthe},{{\bf{\Sigma}}_e})$, respectively. Since the parameters are complex-valued, the FIMs related to the measurements can be calculated based on the following formula \cite{complexFIM1,complexFIM2}: 
\begin{equation}
    {\bf{I}}({\bf{y}}_i;\underline{\bthe})= {\rm{E}}\left\{ \left[ \frac{\partial ln(p_\bthe({\bf{y}}_i))}{\partial \underline{\bthe}^*}\right]  \left[ \frac{\partial ln(p_\bthe({\bf{y}}_i))}{\partial \underline{\bthe}^*}\right]^H \right\} \label{FIMgenFormula}
\end{equation}
for $i \in \{b,e\}$, where $\underline{\bthe}\triangleq[\bthe^T \bthe^H]^T$, $\underline{\bthe}^*$ denotes the conjugate of $\underline{\bthe}$, and $p_\theta({\bf{y}}_b)$ and $p_\theta({\bf{y}}_e)$ are the likelihood functions of the measurements at Bob and Eve, respectively. Based on the expression in \eqref{FIMgenFormula}, the FIMs at Bob and Eve can be obtained as in the following lemma. 

\textit{{\textbf{Lemma~1:}} The FIMs at Bob and Eve can be expressed as}
\begin{align}\label{FIMmainMatrix}
      &{\bf{I}}({\bf{y}}_i;\underline{\bthe})=
      \\\notag
      &\resizebox{8cm}{!}{
       \setlength\arraycolsep{0pt}
      $\begin{bmatrix}
      ({\bf{H}}_{ri}{\bf{\Omega}}{\bf{H}}_{ar}{\bf{P}} )^H
      \boldsymbol{\Sigma}_i^{-1^*}
      ({\bf{H}}_{ri}{\bf{\Omega}}{\bf{H}}_{ar}{\bf{P}} )
      & {\bf{0}} \\
      {\bf{0}} & ({\bf{H}}_{ri}{\bf{\Omega}}{\bf{H}}_{ar}{\bf{P}} )^T
      \boldsymbol{\Sigma}_i^{-1}({\bf{H}}_{ri}{\bf{\Omega}}{\bf{H}}_{ar}{\bf{P}} )^*
      \end{bmatrix}$
      }
\end{align}
\textit{for $i\in\{b,e\}$.}

\textbf{Proof:} The probability density function of a  circularly symmetric complex Gaussian random vector ${\bf{Z}} \sim \mathcal{CN}({\bf{0}},{\bf{\Sigma}}_z)$ can be written as \cite{guassian}
\begin{equation}\label{noisepdf}
    p({\bf{z}})=\frac{1}{\pi^n\deter({\bf{\Sigma}}_z)} e^{- {\bf{z}}^H{\bf{\Sigma}}_z^{-1}{\bf{z}}} 
\end{equation}
where ${\bf{\Sigma}}_z= {\rm{E}}\{{\bf{Z}}{\bf{Z}}^H\}$ and ${\rm{E}}\{{\bf{Z}}{\bf{Z}}^T\}={\boldsymbol{0}}$. By employing (\ref{noisepdf}) for $\Eta_b$ in \eqref{y_bob} and $\Eta_e$ in \eqref{y_eve}, $p_\bthe({\bf{y}}_b)$ and $p_\bthe({\bf{y}}_e)$ can be obtained, and the derivatives required for (\ref{FIMgenFormula}) can be written as follows \cite{complexFIM2}, \cite[eqns.~(81), (82)]{MatrixCookbook}:
\begin{align} \label{NormalDerivative}
    \frac{\partial ln(p_\theta({\bf{y}}_i))}{\partial \bthe} &= \frac{\partial }{\partial \bthe}\left( -({\bf{y}}_i-{\bf{H}}_{i}{\bthe})^H {\bf{\Sigma}}_{n_i}^{-1}({\bf{y}}_i-{\bf{H}}_{i}{\bthe}) \right)\\ \notag
      &=
      \left[-({\bf{y}}_i-{\bf{H}}_{i}{\bthe})^H {\bf{\Sigma}}_{n_i}^{-1} (-{\bf{H}}_{i} )\right]^T \\ \notag 
      &= 
      {\bf{H}}_{i}^T {\bf{\Sigma}}_{n_i}^{-1} ({\bf{y}}_i-{\bf{H}}_{i}{\bthe})^*, 
\end{align}
\begin{align} \label{ConjugateDerivative}
     \frac{\partial ln(p_\theta({\bf{y}}_i))}{\partial \bthe^*} &= 
    \frac{\partial }{\partial \bthe^*}\left( ({\bf{y}}_i-{\bf{H}}_{i}{\bthe})^H {\bf{\Sigma}}_{n_i}^{-1}({\bf{y}}_i-{\bf{H}}_{i}{\bthe}) \right)\\ \notag
      &=  
       {\bf{H}}_{i}^H {\bf{\Sigma}}_{n_i}^{-1} ({\bf{y}}_i-{\bf{H}}_{i}{\bthe} )  
\end{align}
where ${\bf{H}}_{i} \triangleq {\bf{H}}_{ri}{\bf{\Omega}}{\bf{H}}_{ar}{\bf{P}}$.
%\textcolor{red}{Based on equations 81,82 on page 11 of {\bf{matrix cookbook}}, and comparing the results of with the results of page 15 in {\bf{R1}}, I realized that when we take the derivative of Hermitian(or transpose) of a matrix, we don't make any changes at the result of the derivative, but when we take the derivative of a normal matrix, we need to get transpose(not hermitian) of the derivative result.} 
By defining $l_i \triangleq ln(p_\bthe({\bf{y}}_i))$ and inserting the results in (\ref{NormalDerivative}) and (\ref{ConjugateDerivative}) into (\ref{FIMgenFormula}), the following expressions can be obtained:
\begin{align}   \nonumber
    &{\bf{I}}({\bf{y}}_i;\underline{\bthe}) =
    {\rm{E}}\left\{\begin{bmatrix}
        \begin{bmatrix}
            \frac{\partial l_i}{\partial \bthe^*} \\
            \\
            \frac{\partial l_i}{\partial \bthe} 
        \end{bmatrix} &
        \begin{bmatrix}
            \frac{\partial l_i}{\partial \bthe^*} \\
            \\
            \frac{\partial l_i}{\partial \bthe}
        \end{bmatrix}^H
    \end{bmatrix}\right\} \\ \notag
      &=  {\rm{E}}\left\{
      \begin{bmatrix}
        (\frac{\partial l_i}{\partial \bthe^*})
        (\frac{\partial l_i}{\partial \bthe^*})^H &
        (\frac{\partial l_i}{\partial \bthe^*})
        (\frac{\partial l_i}{\partial \bthe})^H \\
        \\
        (\frac{\partial l_i}{\partial \bthe})
        (\frac{\partial l_i}{\partial \bthe^*})^H &
        (\frac{\partial l_i}{\partial \bthe})
        (\frac{\partial l_i}{\partial \bthe})^H 
      \end{bmatrix}
    \right\}\\ \label{FIM_derivatives}
    &= \resizebox{8cm}{!}{
    $\begin{bmatrix}
        {\bf{H}}_{i}^H
        \boldsymbol{\Sigma}_i^{-1}
        {\rm{E}} \left\{\Eta_i\Eta_i^H \right\}
        \boldsymbol{\Sigma}_i^{-1^*}
        {\bf{H}}_{i}  &
        {\bf{H}}_{i}^H
        \boldsymbol{\Sigma}_i^{-1}
        {\rm{E}} \left\{\Eta_i\Eta_i^T \right\}
        \boldsymbol{\Sigma}_i^{-1^*}
        {\bf{H}}_{i}^* \\
        \\
        {\bf{H}}_{i}^T
        \boldsymbol{\Sigma}_i^{-1}
        {\rm{E}} \left\{\Eta_i^*\Eta_i^H \right\}
        \boldsymbol{\Sigma}_i^{-1^*}
        {\bf{H}}_{i}  &
        {\bf{H}}_{i}^T
        \boldsymbol{\Sigma}_i^{-1}
        {\rm{E}} \left\{\Eta_i^*\Eta_i^T \right\}
        \boldsymbol{\Sigma}_i^{-1^*}
        {\bf{H}}_{i}^* 
    \end{bmatrix}$}.
\end{align}
Since $\Eta_b$ and $\Eta_e$ are circularly symmetric complex Gaussian random vectors, it is known that ${\rm{E}}\{\Eta_i\Eta_i^T\}={\boldsymbol{0}}$ and ${\rm{E}}\{\Eta_i\Eta_i^H\}={\bf{\Sigma}}_{i}$. Also, it is noted that ${\rm{E}}\{\Eta_i^*\Eta_i^T\}={\rm{E}}\{(\Eta_i\Eta_i^H)^*\}={\bf{\Sigma}}_i^*$. Thus, the proof is completed by inserting these values into the final expression in \eqref{FIM_derivatives}.\hfill$\blacksquare$

Since the first half of the parameter vector $\underline{\bthe}$ corresponds to the main parameter $\bthe$ of interest, we focus on the trace of the first $k\times k$ block of the FIM in \eqref{FIMmainMatrix} of Lemma~1. Accordingly, the trace of the FIM related to $\bthe$ can be calculated, for $i\in\{b,e\}$, as follows:
\begin{gather}\label{objectivefun_1}
    {\rm{tr}}\{  {\bf{I}}({\bf{y}}_i;{\bthe}) \} =
    {\rm{tr}} \{ ({\bf{H}}_{ri}{\bf{\Omega}}{\bf{H}}_{ar}{\bf{P}} )^H
      \boldsymbol{\Sigma}_i^{-1^*}
      ({\bf{H}}_{ri}{\bf{\Omega}}{\bf{H}}_{ar}{\bf{P}}) \} 
\end{gather}
Since $\boldsymbol{\Sigma}_i$ is positive definite for $i\in\{b,e\}$, the overall matrix in the trace operator in \eqref{objectivefun_1} has real-valued diagonal elements (as it can be expressed as the multiplication of a complex matrix with its Hermitian transpose). Therefore, taking the conjugate of the matrix inside trace operator in \eqref{objectivefun_1} results in the same trace; i.e.,
\begin{gather}\label{objectivefun_2}
    {\rm{tr}}\{  {\bf{I}}({\bf{y}}_i;{\bthe}) \} =
      {\rm{tr}} \{({\bf{H}}_{ri}{\bf{\Omega}}{\bf{H}}_{ar}{\bf{P}} )^T
      \boldsymbol{\Sigma}_i^{-1}({\bf{H}}_{ri}{\bf{\Omega}}{\bf{H}}_{ar}{\bf{P}} )^* \}  
\end{gather}
In addition, since ${\rm{tr}}\{{\bf{A}}\}={\rm{tr}}\{{\bf{A}}^T\}$, \eqref{objectivefun_2} can also be stated as
\begin{gather}\label{objectivefun}
    {\rm{tr}}\{  {\bf{I}}({\bf{y}}_i;{\bthe}) \} =
 {\rm{tr}}\{({\bf{H}}_{ri}{\bf{\Omega}}{\bf{H}}_{ar}{\bf{P}} )^H
        \boldsymbol{\Sigma}_i^{-1}({\bf{H}}_{ri}{\bf{\Omega}}{\bf{H}}_{ar}{\bf{P}} ) \} 
\end{gather}
which is considered as the performance metric to quantify the estimation accuracy.
%The last line is obtained since $tr\{{\bf{A}}\}=tr\{{\bf{A}}^T\}$ and a complex matrix is multiplied by its Hermitian transpose; hence, all the diagonal elements of the argument inside the trace operator are real numbers. Therefore, the $\mathfrak{R}\{.\}$ operator is redundant and eliminated. 

\textbf{Remark~1}: For the proposed optimal RIS design and power allocation problem in \eqref{eq:optProblem} (similarly for \eqref{eq:RISomegaProblem} and \eqref{eq:PAoptProblem}), the transmitter is assumed to know the channel coefficients (i.e., ${\bf{H}}_{ar}$, ${\bf{H}}_{rb}$, and ${\bf{H}}_{re}$), and the covariance matrices ($\boldsymbol{\Sigma}_{b}$ and $\boldsymbol{\Sigma}_{e}$), as can be noted from \eqref{objectivefun}. As the transmitter and the intended receiver are cooperating, information about ${\bf{H}}_{ar}$, ${\bf{H}}_{rb}$, and $\boldsymbol{\Sigma}_{b}$ can be obtained by the transmitter via conventional estimation and feedback procedures. However, for learning ${\bf{H}}_{re}$ and $\boldsymbol{\Sigma}_{e}$, the transmitter is required to have some information about the location of the eavesdropper, the channel model in the environment, and the receiver at the eavesdropper. If ${\bf{H}}_{re}$ and $\boldsymbol{\Sigma}_{e}$ can be learned perfectly, the analysis in this manuscript applies directly. In the case of imperfect information, the inaccuracy about ${\bf{H}}_{re}$ and $\boldsymbol{\Sigma}_{e}$ can be modeled by statistical knowledge, and an averaging operation can be performed as follows: Based on \eqref{objectivefun}, the trace of the FIM can be expressed for the eavesdropper as 
\begin{equation} \label{objectivefunEav}
    {\rm{tr}}\{  {\bf{I}}({\bf{y}}_e;{\bthe}) \} =
    {\rm{tr}}\{
    {\bf{P}}{\bf{H}}_{ar}^H{\bf{\Omega}}^H
    {\bf{E}}
    {\bf{\Omega}}{\bf{H}}_{ar}{\bf{P}}  \} .
\end{equation}
where ${\bf{E}}\triangleq {\bf{H}}_{re}^H\boldsymbol{\Sigma}_e^{-1}{\bf{H}}_{re}$. Suppose that the transmitter does not know ${\bf{E}}$ perfectly but it has statistical information that ${\bf{E}}$ takes $M_e$ possible values with known probabilities. Namely, ${\bf{E}}={\bf{E}}^{(j)}$ with probability $\varsigma^{(j)}$ for $j=1,\ldots,M_e$. Then, the average value of the trace of the FIM can be calculated for the eavesdropper as
\begin{align}
&\sum_{j=1}^{M_e}\varsigma^{(j)}{\rm{tr}}\{
    {\bf{P}}{\bf{H}}_{ar}^H{\bf{\Omega}}^H
    {\bf{E}}^{(j)}
    {\bf{\Omega}}{\bf{H}}_{ar}{\bf{P}}  \}
    \\\label{eq:AvgTrace}
    &=
    {\rm{tr}}\{
    {\bf{P}}{\bf{H}}_{ar}^H{\bf{\Omega}}^H
    \overline{{\bf{E}}}
    {\bf{\Omega}}{\bf{H}}_{ar}{\bf{P}}  \}
\end{align}
where $\overline{{\bf{E}}}\triangleq\sum_{j=1}^{M_e}\varsigma^{(j)}{\bf{E}}^{(j)}$. Since \eqref{eq:AvgTrace} is in the same form as \eqref{objectivefunEav}, the analysis for the case of perfect knowledge applies for the case of imperfect statistical knowledge, as well, by considering the average value of the trace of the FIM as the estimation performance metric.\hfill$\square$

\subsection{RIS Phase Profile Design}\label{sec:RISonly}

In this section, we assume ${\bf{P}}$ to be fixed and aim to solve (\ref{eq:RISomegaProblem}) for deriving the optimal RIS phase profile for maximizing the average Fisher information at Bob under the secrecy constraint. Based on (\ref{objectivefun}), we can reformulate (\ref{eq:RISomegaProblem}) as
%\textcolor{red}{Hocam, here the problem is that our main problem (3) is defined for $I(y;\theta)$ while our proof is for $I(y;\underbar\theta)$ I think we need to mention that since the trace of FIM is a real-valued metric, this change in objective only has caused the multiplier of 2 in our objective and does not affect our optimization parameters. But I don't know to eliminate that coefficient or how to make and discuss this argument.}
\begin{subequations}\label{eq:RISoptProblem}
\begin{align}\label{eq:RISoptProblema}
    \max_{{\bf{\Omega}}}\quad &  {\rm{tr}} \{({\bf{H}}_{rb}{\bf{\Omega}}{\bf{H}}_{ar}{\bf{P}} )^H
        \boldsymbol{\Sigma}_b^{-1}({\bf{H}}_{rb}{\bf{\Omega}}{\bf{H}}_{ar}{\bf{P}} ) \}  \\\label{eq:RISoptProblemb}
    {\rm{s.t.}}\quad & {\rm{tr}} \{({\bf{H}}_{re}{\bf{\Omega}}{\bf{H}}_{ar}{\bf{P}} )^H
        \boldsymbol{\Sigma}_e^{-1}({\bf{H}}_{re}{\bf{\Omega}}{\bf{H}}_{ar}{\bf{P}} ) \}  \leq \Delta  \\\label{eq:RISoptProblemc}
    \quad & \big{|}[{\bf{\Omega}}]_{\ell,\ell}\big{|}=1, ~~\ell=1,\ldots,r
\end{align}
\end{subequations}
%where ${\bf{\Omega}}=diag(\omega_1, \omega_2,\dots,\omega_r)$.
To solve this problem, the following proposition is provided.

\textit{{\textbf{Proposition~1:}} 
Define ${\boldsymbol{\omega}}\in \mathbb{C}^{r}$ as a column vector consisting of the diagonal elements of ${\bf{\Omega}}$; that is, ${\boldsymbol{\omega}}=[\omega_1,\ldots,\omega_r]^T$. Then, we can reformulate (\ref{eq:RISoptProblem}) as follows:}
\begin{subequations}\label{eq:NEWRISoptProblem}
\begin{align} \label{eq:NEWRISoptProblema}
     \max_{{\boldsymbol{\omega}}} &\; \; \;    {\boldsymbol{\omega}}^H{\boldsymbol{\mathcal{Q}}}_{b}{\boldsymbol{\omega}}  \\\label{eq:NEWRISoptProblemb}
      {\rm{s.t.}} ~ &\; \; \;  {\boldsymbol{\omega}}^H{\boldsymbol{\mathcal{Q}}}_{e}{\boldsymbol{\omega}}\   \leq \Delta   \\ \label{eq:NEWRISoptProblemc}
         &\; \; \; |\omega_\ell| = 1, ~~\ell=1,\ldots,r 
\end{align}
\textit{where ${\boldsymbol{\mathcal{Q}}}_i \in \mathbb{C}^{r\times r}$ is a Hermitian matrix that is positive definite for each $i\in\{b,e\}$.}
\end{subequations}

\textbf{Proof:} Considering the expressions in \eqref{eq:RISoptProblema} and \eqref{eq:RISoptProblemb}, we first express the elements of the matrices ${\bf{H}}_{rb}$, ${\bf{H}}_{re}$, and ${\bf{H}}_{ar}{\bf{P}}$ as follows:
\begin{align} \label{Hri}
&{\bf{H}}_{ri} \triangleq 
    \begin{bmatrix}
a_{11} & a_{12} & \dots& a_{1r} \\ 
a_{21} & a_{22} & \dots& a_{2r} \\ 
\vdots &  & \ddots &  \\
 a_{n_i1} &a_{n_i2}  & \dots &a_{n_ir}  \\ 
\end{bmatrix}~{\rm{for}}~i\in\{b,e\}  \\ \label{Har}
&{\bf{H}}_{ar}{\bf{P}} \triangleq 
    \begin{bmatrix}
b_{11} & b_{12} & \dots& b_{1k} \\ 
b_{21} & b_{22} & \dots & b_{2k} \\ 
\vdots &  & \ddots &  \\
b_{r1}  & b_{r2} &\dots& b_{rk}  \\ 
\end{bmatrix}
\end{align}
Accordingly, ${\bf{H}}_{i}$, which was defined as ${\bf{H}}_{i} \triangleq {\bf{H}}_{ri}{\bf{\Omega}}{\bf{H}}_{ar}{\bf{P}}$ in the proof of Lemma~1, can be obtained as 
\begin{align}
    {\bf{H}}_{i} &=
    \resizebox{7cm}{!}{
    $\setlength\arraycolsep{2pt}
    \begin{bmatrix}
    \sum\limits_{j=1}^{r}(a_{1j}\omega_{j}b_{j1}) & \sum\limits_{j=1}^{r}(a_{1j}\omega_{j}b_{j2})  & \dots& &\sum\limits_{j=1}^{r}(a_{1j}\omega_{j}b_{jk})\\ 
\sum\limits_{j=1}^{r}(a_{2j}\omega_{j}b_{j1}) & \sum\limits_{j=1}^{r}(a_{2j}\omega_{j}b_{j2})  & \dots & & \sum\limits_{j=1}^{r}(a_{2j}\omega_{j}b_{jk}) \\ 
\vdots &  & \ddots &  & \vdots \\
\sum\limits_{j=1}^{r}(a_{n_ij}\omega_{j}b_{j1})& \dots & & &\sum\limits_{j=1}^{r}(a_{n_ij}\omega_{j}b_{jk})\\
\end{bmatrix}$
} \\ \label{Hi}
 &= \sum\limits_{j=1}^{r} \omega_j {\bf{S}}_{i,j}
\end{align}
where ${\bf{S}}_{i,j}$ is defined as 
\begin{equation} \label{S_ij}
    {\bf{S}}_{i,j} \triangleq 
    \begin{bmatrix}
        a_{1j}b_{j1} & a_{1j}b_{j2}  & \dots  & a_{1j}b_{jk}  \\ 
        a_{2j}b_{j1} & a_{2j}b_{j2}  & \dots  & a_{2j}b_{jk}  \\ 
        \vdots       &               & \ddots &     \vdots          \\
        a_{n_ij}b_{j1} & a_{n_ij}b_{j2}   & \dots          & a_{n_ij}b_{jk}  \\
    \end{bmatrix}.
\end{equation}
Similarly, ${\bf{H}}_{i}^H$ can be written as ${\bf{H}}_{i}^H=\sum\limits_{j=1}^{r} \omega_j^* {\bf{S}}_{i,j}^H$. Based on the preceding expressions for ${\bf{H}}_{i}$ and ${\bf{H}}_{i}^H$, (\ref{eq:RISoptProblema}) and (\ref{eq:RISoptProblemb}) can be calculated as follows:
\begin{align}\notag
    &{\rm{tr}} \left\{ 
    {\bf{H}}_{i}^H {\boldsymbol{\Sigma}_i^{-1}}{\bf{H}}_{i}
    \right\}  
    \\\notag
    &=  
    {\rm{tr}}\left\{ 
    \left(\sum\limits_{j=1}^{r} \omega_j^*
    {\bf{S}}_{i,j} ^H \right)
    {\boldsymbol\Sigma}_i^{-1}
    \left(\sum\limits_{l=1}^{r} \omega_l
   {\bf{S}}_{i,l}
    \right)
    \right\} \\
     &= \sum\limits_{j=1}^r \sum\limits_{l=1}^r \omega_{j}^* \omega_{l}
    \, {\rm{tr}} \left\{
    {\bf{S}}_{i,j} ^H
    {\boldsymbol\Sigma}_i^{-1}
    {\bf{S}}_{i,l}
    \right\}. \label{XZX}
\end{align}
Next, we define matrix ${\boldsymbol{\mathcal{Q}}}_i \in \mathbb{C}^{r\times r}$ such that its entry in the $j$th row and the $l$th column is given by $[{\boldsymbol{\mathcal{Q}}}_i]_{j,l}={\rm{tr}}\{ {\bf{S}}_{i,j}^H  {\boldsymbol\Sigma}_i^{-1} {\bf{S}}_{i,l}\}$.
%each entry is $q_{i,j,l} \triangleq tr\{ {\bf{S}}_{i,j}^H  {\boldsymbol\Sigma}_i^{-1} {\bf{S}}_{i,l}\}$. Additionally, let ${\boldsymbol{\omega}}=[\omega_1, \omega_2, \dots, \omega_r]^T \in \mathbb{C}^{r} $. 
Then, it is noted that the expression in \eqref{XZX} can be stated as ${\boldsymbol{\omega}}^H{\boldsymbol{\mathcal{Q}}}_{i}{\boldsymbol{\omega}}$ with ${\boldsymbol{\omega}}=[\omega_1, \omega_2, \dots, \omega_r]^T$. Therefore, (\ref{eq:RISoptProblema}) and (\ref{eq:RISoptProblemb}) are equal to (\ref{eq:NEWRISoptProblema}) and (\ref{eq:NEWRISoptProblemb}), respectively, as claimed in the proposition. In addition, the elements of ${\boldsymbol{\mathcal{Q}}}_i$ satisfy the following relations:
\begin{align}\notag
[{\boldsymbol{\mathcal{Q}}}_i]_{j,l}^* &= {\rm{tr}}\{ {\bf{S}}_{i,j}^H  {\boldsymbol\Sigma}_i^{-1} {\bf{S}}_{i,l} \}^* \\ \notag
         &= {\rm{tr}}\{{\bf{S}}_{i,j}^T ({\boldsymbol\Sigma}_i^{-1})^*  {\bf{S}}_{i,l}^*\} \\ \notag
          &= {\rm{tr}}\{({\bf{S}}_{i,j}^T ({\boldsymbol\Sigma}_i^{-1})^*  {\bf{S}}_{i,l}^*)^T\} \\ \notag
          &= {\rm{tr}}\{{\bf{S}}_{i,l}^H ({\boldsymbol\Sigma}_i^{-1})^H  {\bf{S}}_{i,j}\} \\ \notag
          &= {\rm{tr}}\{{\bf{S}}_{i,l}^H {\boldsymbol\Sigma}_i^{-1}  {\bf{S}}_{i,j}\} \\ \label{eq:HermitianProof}
          &= [{\boldsymbol{\mathcal{Q}}}_i]_{l,j}.
\end{align}
where the third equality is based on the property that ${\rm{tr}}\{{\bf{A}}\}={\rm{tr}}\{{\bf{A}}^T\}$ and the fifth equality is due to the fact that the covariance matrix of a complex random vector is a Hermitian matrix (so is its inverse). Overall, it is shown in \eqref{eq:HermitianProof} that $[{\boldsymbol{\mathcal{Q}}}_i]_{j,l}^*=[{\boldsymbol{\mathcal{Q}}}_i]_{l,j}$ for any $l$ and $j$, implying that ${\boldsymbol{\mathcal{Q}}}_i$ is a Hermitian matrix.

%\textcolor{blue}{To show that ${\boldsymbol{\mathcal{Q}}}_i$ is hermitian, we need to show that $[{\boldsymbol{\mathcal{Q}}}^H_i]_{j,l}=[
%{\boldsymbol{\mathcal{Q}}}_i]_{j,l} $ for any i and j. Hence, we can write
%\begin{align} \notag
%  [{\boldsymbol{\mathcal{Q}}}^H_i]_{j,l} &=[{\boldsymbol{\mathcal{Q}}}_i]^*_{l,j} \\ \notag
%  &= tr\{{\bf{S}}_{i,l}^H ({\boldsymbol\Sigma}_i^{-1})  {\bf{S}}_{i,j}\}^* \\ \notag
%  &= tr\{{\bf{S}}_{i,l}^T ({\boldsymbol\Sigma}_i^{-1})^*  {\bf{S}}_{i,j}^*\} \\ \notag
%  &= tr\{{\bf{S}}_{i,l}^T ({\boldsymbol\Sigma}_i^{-1}) ^* {\bf{S}}_{i,j}^*\}^T \\ \notag
%  &= tr\{{\bf{S}}_{i,j}^H ({\boldsymbol\Sigma}_i^{-1})^H  {\bf{S}}_{i,l}\} \\ \notag 
%  &= tr\{{\bf{S}}_{i,j}^H ({\boldsymbol\Sigma}_i^{-1})  {\bf{S}}_{i,l}\} \\ 
%  &= [{\boldsymbol{\mathcal{Q}}}_i]_{j,l}
%\end{align}
%}

To prove the positive definiteness of ${\boldsymbol{\mathcal{Q}}}_i$, consider a generic non-zero complex vector ${\bf{z}}\in \mathbb{C}^r$. Then, the following relations can be obtained (cf.~\eqref{XZX}):
\begin{align}\notag
{\bf{z}}^H{\boldsymbol{\mathcal{Q}}}_i{\bf{z}}  
%=& \sum_{j=1}^{r}%\sum_{l=1}^{r}z_j^*z_l[{\boldsymbol{\mathcal{Q}}}_i]_{j,l}
%\\
&= \sum_{j=1}^{r}\sum_{l=1}^{r}z_j^*z_l\,{\rm{tr}}\{ {\bf{S}}_{i,j}^H  {\boldsymbol\Sigma}_i^{-1} {\bf{S}}_{i,l}\}
\\\notag
&=
    {\rm{tr}}\left\{ 
    \left(\sum\limits_{j=1}^{r} z_j^*
    {\bf{S}}_{i,j} ^H \right)
    {\boldsymbol\Sigma}_i^{-1}
    \left(\sum\limits_{l=1}^{r} z_l
   {\bf{S}}_{i,l}
    \right)
    \right\} \\\label{PSD}
&\triangleq
{\rm{tr}} \left\{ 
{\bf{X}}_i^H {\boldsymbol{\Sigma}_i^{-1}}{\bf{X}}_i
\right\} >0
%&= \sum\limits_{j=1}^r {\bf{x}}_j^H{\boldsymbol\Sigma}_i^{-1} %{\bf{x}}_j \notag\\ 
%&>  0, \notag
\end{align}
where ${\bf{X}}_i$ is defined as ${\bf{X}}_i\triangleq \sum_{l=1}^{r} z_l {\bf{S}}_{i,l}$. The last expression in \eqref{PSD} is positive due to the positive definiteness of ${\boldsymbol{\Sigma}}_i$. (In particular, if ${\bf{x}}_{i,j}$ denotes the $j$th column of ${\bf{X}}_i$, \eqref{PSD} becomes $\sum_j {\bf{x}}_{i,j}^H{\boldsymbol\Sigma}_i^{-1}{\bf{x}}_{i,j}$, which is always positive.) Hence, ${\boldsymbol{\mathcal{Q}}}_i$ is a positive definite matrix for each $i\in\{b,e\}$.\hfill$\blacksquare$

\textbf{Remark~2}: Since the expressions in (\ref{eq:NEWRISoptProblema}) and (\ref{eq:NEWRISoptProblemb}) correspond to the traces of the FIMs (see \eqref{objectivefun} and \eqref{eq:RISoptProblem}), they should be non-negative for any ${\boldsymbol{\omega}}$ by definition. Therefore, ${\boldsymbol{\mathcal{Q}}}_b$ in (\ref{eq:NEWRISoptProblema}) and ${\boldsymbol{\mathcal{Q}}}_e$ in (\ref{eq:NEWRISoptProblemb}) being positive definite matrices is in compliance with this intuition.\hfill$\square$

\textbf{Remark~3}: Since ${\boldsymbol{\mathcal{Q}}}_i$ is a Hermitian matrix for each $i\in\{b,e\}$, its eigenvalues are real-valued and \eqref{eq:NEWRISoptProblema} (similarly, \eqref{eq:NEWRISoptProblemb}) can be upper and lower bounded as follows \cite{EigBound}:
\begin{equation}\label{eq:LBUB}
    \lambda_{i,\min}\,{\boldsymbol{\omega}}^H{\boldsymbol{\omega}} \leq{\boldsymbol{\omega}}^H{\boldsymbol{\mathcal{Q}}}_{i}{\boldsymbol{\omega}}\leq \lambda_{i,\max}\,{\boldsymbol{\omega}}^H{\boldsymbol{\omega}}
\end{equation}
where $\lambda_{i,\min}$ and $\lambda_{i,\max}$ denote, respectively, the minimum and maximum eigenvalues of ${\boldsymbol{\mathcal{Q}}}_{i}$. Since ${\boldsymbol{\omega}}^H{\boldsymbol{\omega}}=r$ due to the constraint in \eqref{eq:NEWRISoptProblemc}, $r\lambda_{i,\min}\leq{\boldsymbol{\omega}}^H{\boldsymbol{\mathcal{Q}}}_{i}{\boldsymbol{\omega}}\leq r\lambda_{i,\max}$ can be obtained. However, $r\lambda_{i,\min}$ and $r\lambda_{i,\max}$ may not be tight bounds for ${\boldsymbol{\omega}}^H{\boldsymbol{\mathcal{Q}}}_{i}{\boldsymbol{\omega}}$ in \eqref{eq:NEWRISoptProblem} since the eigenvectors corresponding to the minimum and maximum eigenvalues of ${\boldsymbol{\mathcal{Q}}}_{i}$ may not be multiples of the all-ones vector.\hfill$\square$

%\textbf{Proof}: 
% One can show that for any Hermitian matrix, eigenvalues are real-valued. 
%With eigenvalue decomposition, we can rewrite 
%${\boldsymbol{\mathcal{Q}}}_i= \sum\limits_{k=1}^{n_i} \lambda_{i,k} {\bf{v}}_{i,k} {\bf{v}}_{i,k}^H$, where $\lambda_{i,k}$ and ${\bf{v}}_{i,k}$ are eigenvalue and eigenvectors, respectively. Therefore, for any $\textbf{x}\in \mathbb{C}^r$, \textcolor{red}{where $|\textbf{x}|=\textbf{1}$} we can write 
%\begin{subequations}
%\begin{align}
%    \textbf{x}^H {\boldsymbol{\mathcal{Q}}}_i \textbf{x} &= %\sum\limits_{k=1}^{n_i} \lambda_{i,k}  \textbf{x}^H %{\bf{v}}_{i,k} {\bf{v}}_{i,k}^H \textbf{x} \\
%    &\leq \lambda_{i,max} \textbf{x}^H \left(\sum\limits_{k=1}^{n_i}  {\bf{v}}_{i,k} {\bf{v}}_{i,k}^H \right) \textbf{x} \\
%     &=  \lambda_{i,max}  \textbf{x}^H  \textbf{x}\\
%     &\leq  r \lambda_{i,max}.       
%\end{align}
%\end{subequations}
%similarly we can show that $\lambda_{i,min}$ is the lower bound %and the proof is completed.

Based on Proposition~1, the RIS phase profile design problem in \eqref{eq:RISoptProblem} is stated in a more explicit form as in \eqref{eq:NEWRISoptProblem}. The objective function in \eqref{eq:NEWRISoptProblema} and the first constraint in \eqref{eq:NEWRISoptProblemb} involve quadratic functions of the optimization variables. However, due to the constraints in (\ref{eq:NEWRISoptProblemc}), the RIS phase profile optimization becomes a non-convex problem, namely, non-convex quadratic constrained quadratic programming (QCQP). As in \cite{SDR_QCQP}, this problem can be solved via the SDR approach. Since ${\boldsymbol{\omega}}^H{\boldsymbol{\mathcal{Q}}}_{i}{\boldsymbol{\omega}} ={\rm{tr}}\{{\boldsymbol{\omega}}^H{\boldsymbol{\mathcal{Q}}}_{i}{\boldsymbol{\omega}} \}={\rm{tr}}\{{\boldsymbol{\mathcal{Q}}}_{i}{\boldsymbol{\omega}{\boldsymbol{\omega}}^H} \} $, the problem in \eqref{eq:NEWRISoptProblem} can be reformulated as
\begin{subequations}\label{eq:SDRRISoptProblem}
\begin{align} \label{eq:SDRRISoptProblema}
     \max_{\boldsymbol{W}} &\; \; \;    {\rm{tr}}\{ {\boldsymbol{\mathcal{Q}}}_{b}{\boldsymbol{W}} \} \\\label{eq:SDRRISoptProblemb}
      {\rm{s.t.}} &\; \; \;  {\rm{tr}}\{ {\boldsymbol{\mathcal{Q}}}_{e}{\boldsymbol{W}}\}   \leq \Delta   \\ \label{eq:SDRRISoptProblemc}
      &\; \; \; [{\boldsymbol{W}}]_{ii}=1 \;\;\; {\rm{for}} \;\;\; i=1,\ldots,r \\
      \label{eq:SDRRISoptProblemd}
         &\; \; \; {\boldsymbol{W}} \succeq {\bf{0}} \; , \; {\rm{rank}}({\boldsymbol{W}})=1
\end{align}
\end{subequations}
where ${\boldsymbol{W}} \triangleq \boldsymbol{\omega}\boldsymbol{\omega}^H$. After relaxing the rank constraint, the following convex formulation is obtained:
\begin{subequations}\label{eq:SDRRISoptProblem_2}
\begin{align} \label{eq:SDRRISoptProblema_2}
     \max_{\boldsymbol{W}} &\; \; \;    {\rm{tr}}\{ {\boldsymbol{\mathcal{Q}}}_{b}{\boldsymbol{W}} \} \\\label{eq:SDRRISoptProblemb_2}
      {\rm{s.t.}} &\; \; \;  {\rm{tr}}\{ {\boldsymbol{\mathcal{Q}}}_{e}{\boldsymbol{W}}\}   \leq \Delta   \\ \label{eq:SDRRISoptProblemc_2}
      &\; \; \; [{\boldsymbol{W}}]_{ii}=1 \;\;\; {\rm{for}} \;\;\; i=1,\ldots,r \\
      \label{eq:SDRRISoptProblemd_2}
         &\; \; \; {\boldsymbol{W}} \succeq {\bf{0}} 
\end{align}
\end{subequations}
As stated in \cite{SDR_QCQP}, the problem in \eqref{eq:SDRRISoptProblem_2} can be solved via various approaches such as the interior point methods \cite{boyd_convex}, first-order methods \cite[Chapter~23]{Penalty_projection}, and sequential quadratic programming (SQP) \cite{SQP}. MATLAB toolboxes such as CVX \cite{cvx1,cvx2} and YALMIP \cite{YALMIP1} are two powerful and well-known frameworks based on interior point methods for solving this sort of problems. The solution based on the SDR approach, i.e., the solution of \eqref{eq:SDRRISoptProblem_2}, does not necessarily satisfy the rank-one constraint in \eqref{eq:SDRRISoptProblem}. 
%(If the solution has rank one, then it is optimal.) 
Therefore, various approaches, such as eigenvalue decomposition (ED) and Gaussian randomization (GR), are employed to derive a sub-optimal rank-one solution \cite{ED_GR}. In the ED method, the eigenvector corresponding to the maximum eigenvalue of the solution of \eqref{eq:SDRRISoptProblem_2} is considered as the sub-optimal solution of \eqref{eq:SDRRISoptProblem}. In the GR method, after obtaining the eigenvalue decomposition, $\boldsymbol{W}^*=\boldsymbol{U}\boldsymbol{\Lambda}\boldsymbol{U}^H$, random vectors are generated as $\tilde{\boldsymbol{\omega}}=\boldsymbol{U}\boldsymbol{\Lambda}^{\frac{1}{2}}\mathbf{g}$, where ${\mathbf{g}}\sim \mathcal {CN}(\mathbf {0},\mathbf {I}_{r})$. Then, $\hat{\boldsymbol{\omega}}=e^{j\angle \tilde{\boldsymbol{\omega}}}$ is considered as the sub-optimal solution of \eqref{eq:SDRRISoptProblem_2}. The preceding sub-optimal rank-one solutions do not necessarily satisfy the secrecy constraint in \eqref{eq:NEWRISoptProblemb} or \eqref{eq:SDRRISoptProblemb}. Therefore, we combine them as follows by considering the secrecy constraint: First, the sub-optimal rank-one solution is obtained via the ED method, and the corresponding secrecy constraint is checked. If the secrecy constraint is satisfied, that rank-one solution is declared as the sub-optimal solution of \eqref{eq:NEWRISoptProblem}. Otherwise, rank-one sub-optimal solutions are generated based on the GR method until the secrecy constraint is satisfied, where the random nature of the GR method is utilized to get a feasible solution. 
%This approach is shown in Algorithm~1.

\textbf{Remark~4}: Another approach to deal with rank-one relaxation is to use a penalty method \cite{penaltyMethod1,penaltyMethod2}. In this approach, %since the objective function is non-negative, 
a penalty component can be added to the objective function to obtain a feasible solution. For instance, \cite{penaltyMethod1} uses a factor of the difference between the nuclear norm and the spectral norm of the optimization variable (matrix) as the penalty term. 
%$\rho(||\Omega||_* - ||\Omega||_2)$ as a penalty term, where $\rho$ is a design parameter, and $||\Omega||_*=\sum_m \lambda_m(\Omega)$ and $||\Omega||_2=\lambda_1(\Omega)$ denote the nuclear norm and the spectral norm, respectively. 
Furthermore, Taylor series expansion is used to obtain a convex
upper bound for the penalty term and CVX in MATLAB is employed to solve the resulting problem. In some set-ups where the Gaussian randomization technique has difficulty in obtaining a rank-one solution in a limited amount of time, the penalty approach can be helpful in deriving a feasible solution.\hfill$\square$

As stated in Remark~3, there exists a certain lower bound for ${\boldsymbol{\omega}}^H{\boldsymbol{\mathcal{Q}}}_{e}{\boldsymbol{\omega}}$. Hence, if the secrecy limit $\Delta$ is lower than that bound, \eqref{eq:NEWRISoptProblem} becomes infeasible. For example, \eqref{eq:NEWRISoptProblem} is infeasible if $\Delta<r\lambda_{e,\min}$, where $\lambda_{e,\min}$ denotes the minimum eigenvalue of ${\boldsymbol{\mathcal{Q}}}_{e}$. However, as stated in Remark~3, $r\lambda_{e,\min}$ is not a tight bound in general. To specify the feasibility region for \eqref{eq:NEWRISoptProblem} more accurately, the following problem can be defined:
\begin{subequations}\label{eq:FeasRISoptProblem}
\begin{align} \label{eq:FeasRISoptProblema}
     \min_{\{\omega_i\}_{i=1}^r} &\; \; \;    {\boldsymbol{\omega}}^H{\boldsymbol{\mathcal{Q}}}_{e}{\boldsymbol{\omega}}  \\\label{eq:FeasRISoptProblemb}
      {\rm{s.t.}}~ &~~|\omega_i| = 1  \;\;\; {\rm{for}} \;\;\; i=1,2,\dots,r 
\end{align}
\end{subequations}
Similar to \eqref{eq:NEWRISoptProblem}, this problem can also be solved via the SDR approach to derive a sub-optimal solution. If the minimum value of \eqref{eq:FeasRISoptProblema} corresponding to the sub-optimal solution is denoted by $\Delta _{\min}$, the problem in \eqref{eq:NEWRISoptProblem} becomes feasible for all $\Delta \geq\Delta _{\min}$. 

%\textcolor{red}{Is it necessary to write complexity order?}

\subsection{Optimal Power Allocation}\label{sec:PA_only}

In this section, we assume $\bf{\Omega}$ to be fixed and aim to solve \eqref{eq:PAoptProblem} for deriving the optimal power allocation matrix. Based on \eqref{objectivefun}, we can rewrite \eqref{eq:PAoptProblem} as
\begin{subequations}\label{eq:NEWPAoptProblem}
\begin{align}\label{eq:NEWoptProblema}
    \max_{{\bf{P}}}\quad &{\rm{tr}}\{{\bf{P}}^H{\boldsymbol{A}}_{b}{\bf{P}}\}\\\label{eq:NEWPAoptProblemb}
    {\rm{s.t.}}\quad & {\rm{tr}}\{{\bf{P}}^H{\boldsymbol{A}}_{e}{\bf{P}}\} \leq \Delta  \\\label{eq:NEWPAoptProblemc}
    \quad &{\rm{tr}}\{{\bf{P}}{\bf{P}}^T\}\leq P_\Sigma\\\label{eq:NEWPAoptProblemd}
    \quad & {\bf{P}}{{\bf{P}}^T}\succeq {\bf{0}}.
\end{align}
\end{subequations}
where ${\boldsymbol{A}}_i$ is defined as ${\boldsymbol{A}}_i \triangleq ({\bf{H}}_{ri}{\bf{\Omega}}{\bf{H}}_{ar})^H \boldsymbol{\Sigma}_i^{-1}({\bf{H}}_{ri}{\bf{\Omega}}{\bf{H}}_{ar})$ for $i\in\{b,e\}$. Since $\bf{P}$ is a real diagonal matrix represented by ${\bf{P}}=\diag\{\sqrt{p}_1,\sqrt{p}_2,\ldots,\sqrt{p}_k\}$, \eqref{eq:NEWPAoptProblem} is equivalent to 
\begin{subequations}\label{eq:LP}
\begin{align}\label{eq:LPa}
    \max_{\{{p_i}\}_{i=1}^k}\quad &\sum\limits_{i=1}^k \alpha_i p_i\\ \label{LPb}
    {\rm{s.t.}}\quad & \sum\limits_{i=1}^k \beta_i p_i\leq \Delta  \\\label{LPc}
    \quad &\sum\limits_{i=1}^k  p_i\leq P_\Sigma\\\label{LPd}
    \quad & ~{p}_{i}\geq 0 \;\;\; {\rm{for}} \;\;\; i=1,2,\dots,k 
\end{align}
\end{subequations}
where $\alpha_i \triangleq [{\boldsymbol{A}}_{b}]_{ii}$ and $\beta_i \triangleq [{\boldsymbol{A}}_{e}]_{ii}$. 
The problem in \eqref{eq:LP} corresponds to a standard linear programming (LP) formulation, which can be solved via conventional algorithms for LP \cite{LP1}. (The linear programming solver in MATLAB \cite{LP2}, `linprog', is used in this study for solving \eqref{eq:LP}.)

\subsection{Joint Optimization of RIS Phase Profile and Power Allocation}

This section presents the proposed alternating optimization approach for solving the joint RIS phase profile design and power allocation problem in \eqref{eq:optProblem}. Namely, a solution to \eqref{eq:optProblem} is obtained by iteratively solving the RIS phase design problem in \eqref{eq:SDRRISoptProblem} and the power allocation problem in \eqref{eq:LP}. The details are shown in Algorithm~1. %The initial values for the RIS phase profile and the power allocation matrix are set as follows: 
In this algorithm, the RIS phase profile is initially set to a random phase profile, where the elements have uniformly distributed i.i.d. random phases between $0$ and $2\pi$. 
%Meanwhile, the power allocation matrix is considered to start with equal power allocation initially. 
Then, the power allocation matrix is optimized based on \eqref{eq:LP} considering the RIS phase profile to be fixed, and then the RIS phase profile is updated by solving \eqref{eq:SDRRISoptProblem_2} for the power allocation matrix obtained in the previous step and by applying rank reduction to obtain a rank-one solution. These iterations continue until the objective function does not change significantly, which is determined by parameter $\epsilon$ in Algorithm~1, or until the number $n$ of iterations reaches the maximum number of iterations, $n_{\max}$. (The values of the RIS phase profile and the power allocation matrix at iteration $n$ are denoted by  $\boldsymbol{\omega}^{(n)}$ and $\boldsymbol{P}^{(n)}$, respectively, and ${\boldsymbol{W}}^{(n)}$ denotes the solution of \eqref{eq:SDRRISoptProblem_2} in the $n$th iteration.) 
%In the following algorithm, the error tolerance of ${\epsilon}$ is considered as the stopping criteria for choosing RIS and P, and $n$ is the counter for updating the parameters at each iteration. 
At the end of the iterations, the final values of the RIS phase profile and the power allocation matrix are considered as a solution of \eqref{eq:optProblem}, which are denoted by $\boldsymbol{\Omega}^*$ and ${\bf{P}}^*$ in Algorithm~1.

\begin{algorithm}
\begin{small}
	\caption{Joint optimization of $\boldsymbol{\Omega}$ and       ${\bf{P}}$ for solving \eqref{eq:optProblem}.}
	\begin{algorithmic}[1]
    	\State \textbf{Define:}  convergence tolerance $\epsilon$, the maximum number of iterations $n_{\max}$, iteration index $n$, $1\times n_{\max}$ vector $\textbf{f}_{b}$ for saving the average Fisher information at Bob in each iteration
	\State \textbf{Initialization:}
        \State $n=1$
        \State $\phi_i\leftarrow \mathcal{U}[0,2\pi)$ i.i.d uniform r.v.'s for $i=1,2,\dots,r$
        \State ${\boldsymbol{\omega}}^{(0)}=[ e^{j2\pi {\phi_1}}, e^{j2\pi {\phi_2}},\ldots,e^{j2\pi {\phi_r}}]^T$
        \State \textbf{First iteration:}
        \State ${\boldsymbol{P}}^{(1)} \xleftarrow{}$ 
        {\textrm{solution of \eqref{eq:LP} for fixed ${\boldsymbol{\omega}}^{(0)}$}}
        \State ${\boldsymbol{W}}^{(1)} \xleftarrow{}$ 
        {\textrm{solution of \eqref{eq:SDRRISoptProblem_2} for fixed ${\boldsymbol{P}}^{(1)}$}}
        \State ${\boldsymbol{\omega}}^{(1)} \xleftarrow{}$ {\textrm{rank-one solution obtained from ${\boldsymbol{W}}^{(1)}$ via Algorithm~2}}
        \State ${\bf{f}}_{b}(1) \xleftarrow{}$ {\textrm{value of \eqref{eq:NEWRISoptProblema} for ${\boldsymbol{\omega}}^{(1)}$ and ${\boldsymbol{P}}^{(1)}$}}
        \Repeat
        \State ${\boldsymbol{P}}^{(n+1)} \xleftarrow{}$ \textrm{solution of \eqref{eq:LP} for fixed ${\boldsymbol{\omega}}^{(n)}$}\;
        \State  ${\boldsymbol{W}}^{(n+1)} \xleftarrow{}$ {\textrm{solution of \eqref{eq:SDRRISoptProblem_2} for fixed ${\boldsymbol{P}}^{(n+1)}$}}
        \State run Algorithm~2 based on ${\boldsymbol{W}}^{(n+1)}$
        \If {Algorithm~2 returns a feasible solution}
        \State  ${\boldsymbol{\omega}}^{(n+1)} \xleftarrow{}$ \textrm{rank-one solution obtained from Algorithm~2}
        \Else
        \State $\boldsymbol{\omega}^{(n+1)} \leftarrow \boldsymbol{\omega}^{(n)}$ 
        \EndIf
        \State ${\bf{f}}_{b}(n+1) \xleftarrow{}$ {\textrm{value of \eqref{eq:NEWRISoptProblema} for        ${\boldsymbol{\omega}}^{(n+1)}$ and ${\boldsymbol{P}}^{(n+1)}$}}  
        \State $n=n+1$
        \Until {${\bf{f}}_{b}(n)-\textbf{f}_{b}(n-1) \leq {\epsilon}$ {\rm{or}} $n =  n_{\max}$} 
        \If {$n= n_{\max}$} 
        \State $n^{*}= \max_{{n}}{{\bf{f}}_{b}(n)}$
        \State $\boldsymbol{\Omega}^*=\diag\{{\boldsymbol{\omega}}^{(n^{*})}\}$
         \State ${\bf{P}}^*={\boldsymbol{P}}^{(n^{*})}$ 
        \Else
         \State $\boldsymbol{\Omega}^*=\diag\{{\boldsymbol{\omega}}^{(n)}\}$
         \State ${\bf{P}}^*={\boldsymbol{P}}^{(n)}$ 
         \EndIf
	\end{algorithmic} 
\end{small} 
\end{algorithm}  

In Algorithm~1, the rank reduction approach discussed in Section~\ref{sec:RISonly} is utilized to obtain a rank-one solution for the RIS phase profile based on the solution of \eqref{eq:SDRRISoptProblem_2}, which is shown in Algorithm~2. Since a complex Gaussian random variable is utilized to obtain a solution in the GR part of Algorithm 2, it is possible for the average Fisher information at Bob and Eve to have some fluctuation between certain values. The resulting objective values are saved in a vector named ${\boldsymbol{f}}_{b}$ at each iteration of Algorithm~1, and when the iteration number reaches $n_{\max}$, the value of $\boldsymbol{\omega}$ calculated in the iteration that achieves the maximum value of $\boldsymbol{f}_{b}$ is chosen as the optimal solution.
%Also, $\boldsymbol{\omega}$ and $\textbf{p}$ are the vectors that include diagonal elements of $\boldsymbol{\Omega}$ and $\textbf{P}$, respectively, and $\boldsymbol{\omega}^*$ and $\textbf{p}^*$ are considered as their optimal values.

During the alternating optimization iterations in Algorithm~1, when optimizing the RIS phase profile for a fixed power allocation matrix, the existence of a feasible solution is not guaranteed (please see the explanations at the end of Section~\ref{sec:RISonly}). However, the optimal power allocation problem for a given RIS phase profile is always feasible (please see \eqref{eq:LP}). In Algorithm~1, the feasibility issue is addressed by utilizing the power allocation algorithm and the solutions obtained in the previous iterations. Specifically, when the GR technique in Algorithm~2 reaches the maximum iteration number $L$, due to the infeasibility of the RIS phase profile optimization problem, the solution of the previous iteration in Algorithm~1 is used to derive the rank-one solution for the RIS phase profile.

%-------------------------
\begin{algorithm}
\begin{small} 
	\caption{Rank Reduction via ED and GR}
	\begin{algorithmic}[1]
        \State \textbf{Define:} randomization index $l$, maximum number of randomization operations $L > 1$
        \State \textbf{Given:} Solution of \eqref{eq:SDRRISoptProblem_2} denoted by $\boldsymbol{W}^*$ 
        %rank-one solution from previous iteration $\boldsymbol{\omega}^*_{\rm{prev}}$   
        \State \textbf{Initialization:} $l=1$
        \State \textbf{ED:} \{ 
        $\boldsymbol{W}^*=\boldsymbol{U}\boldsymbol{\Lambda}\boldsymbol{U}^H$
        \State  $\boldsymbol{\Lambda}={\rm{diag}}\{ \lambda_{1}, \lambda_{2}, \dots, \lambda_{r}\}$, where $\lambda_{i}$ is the $i^{th}$ eigenvalue of $\boldsymbol{W}^*$
       % \State sorting $\lambda_{1}\geq \lambda_{2}\geq \dots \geq \lambda_{r}$
        \State Find the maximum eigenvalue $\lambda_{\max}$ of $\boldsymbol{W}^*$ and the corresponding eigenvector $\boldsymbol{v}_{\max}$
        \State $\boldsymbol{\omega}\leftarrow \boldsymbol{v}_{\max}$ \}
         \If {(\ref{eq:NEWRISoptProblemb}) holds for $\boldsymbol{\omega}$}
        \State $\boldsymbol{\omega}^{*} \leftarrow  \boldsymbol{\omega}$ 
        \Else 
        \State \textbf{GR:} \{
        \Repeat
        \State ${\mathbf{g}}\sim \mathcal {CN}(\mathbf {0},\mathbf {I}_{r})$
        \State $\boldsymbol{\omega} \leftarrow \boldsymbol{U}\boldsymbol{\Lambda}^{\frac{1}{2}}\mathbf{g}$
        \State $l=l+1$
        \Until ({\ref{eq:NEWRISoptProblemb}) holds {\rm{or}} $l  =  L$} 
        %\State $\boldsymbol{\omega}^* \leftarrow \boldsymbol{\omega}$ \}
            \If {$l <  L$}
            \State $\boldsymbol{\omega}^{*} \leftarrow \boldsymbol{\omega}$
            \Else
             \State \textrm{no feasible solution is available}
             % k\boldsymbol{v}_{\max}??? or from prev. iteration...
             \EndIf\}
         \EndIf
	\end{algorithmic} 
\end{small}  
\end{algorithm}

\section{Numerical Results}\label{sec:Nume}

In this section, we investigate the theoretical results and the proposed approach based on various numerical examples. As adopted in \cite{TsP2022}, the real and imaginary parts of the channel matrices  ${\bf{H}}_{ar}$, ${\bf{H}}_{rb}$, and ${\bf{H}}_{re}$ in \eqref{y_bob} and \eqref{y_eve} are generated as i.i.d. uniform random variables ranging over $[-0.1$, $0.1]$ by utilizing a single realization in MATLAB with seed $1$. Also, the additive noise vectors in Bob and Eve ($\Eta_b$ in \eqref{y_bob} and $\Eta_e$ in \eqref{y_eve}, respectively) are modeled as independent zero-mean circularly-symmetric Gaussian random vectors with i.i.d. components, where the variance of each component is set to $10^{-5}$. In addition, the observation sizes at Bob and Eve are considered to be $n_b=n_e=2k$, where $k$ denotes the number of parameters \cite{TsP2022}. Moreover, the size of the RIS is taken as $r=36$ $(6\times 6)$ or $r=64$ ($8 \times 8$) in the simulations. Finally, the convergence metric in Algorithm~1 is set to $\epsilon=10^{-1}$.

For each case, we compare our proposed approach of alternating optimization (AO) with two benchmarks referred to as ``RIS only", ``power allocation (PA) only". In the RIS only algorithm, we set the power allocation matrix $\bf{P}$ to be an equal-power allocation matrix, namely, ${\bf{P}} = \diag \{ \sqrt{P_{\Sigma}/{k}},\sqrt{P_{\Sigma}/{k}}, \ldots,\sqrt{P_{\Sigma}/{k}}\}$. %where $p_i={\frac{P_{\Sigma}}{k}}$.
The aim of this case is to investigate the effects of optimizing the RIS phase profile with the unit modulus constraint for secure estimation of an unknown deterministic complex parameter vector by solving \eqref{eq:RISomegaProblem}. In the PA only algorithm, a fixed RIS phase profile ${\bf{\Omega}}= \diag \{e^{j\phi_{1}},e^{j\phi_{2}}\ldots,e^{j\phi_{r}}\}$ 
%${\boldsymbol{\phi}}=[\phi_{1},\phi_{2},\dots,\phi_{r}]$ 
is considered to be generated randomly with $\phi_i$'s being i.i.d. uniform random variables over $[0$,$2\pi)$. The aim of this case is to analyze the effects of secure power allocation with the total power constraint, similar to \cite{TsP2022} but for a different optimization metric of average Fisher information, by solving \eqref{eq:PAoptProblem}. We compare these benchmarks with the joint optimization of the RIS phase profile and power allocation via the proposed AO approach. %It is expected to have better performance compared to both of the earlier scenarios. 

\begin{figure}%[htp]
\vspace{-0.4cm}

\centering

\subfloat{%
  \includegraphics[width=0.9\columnwidth,draft=false]{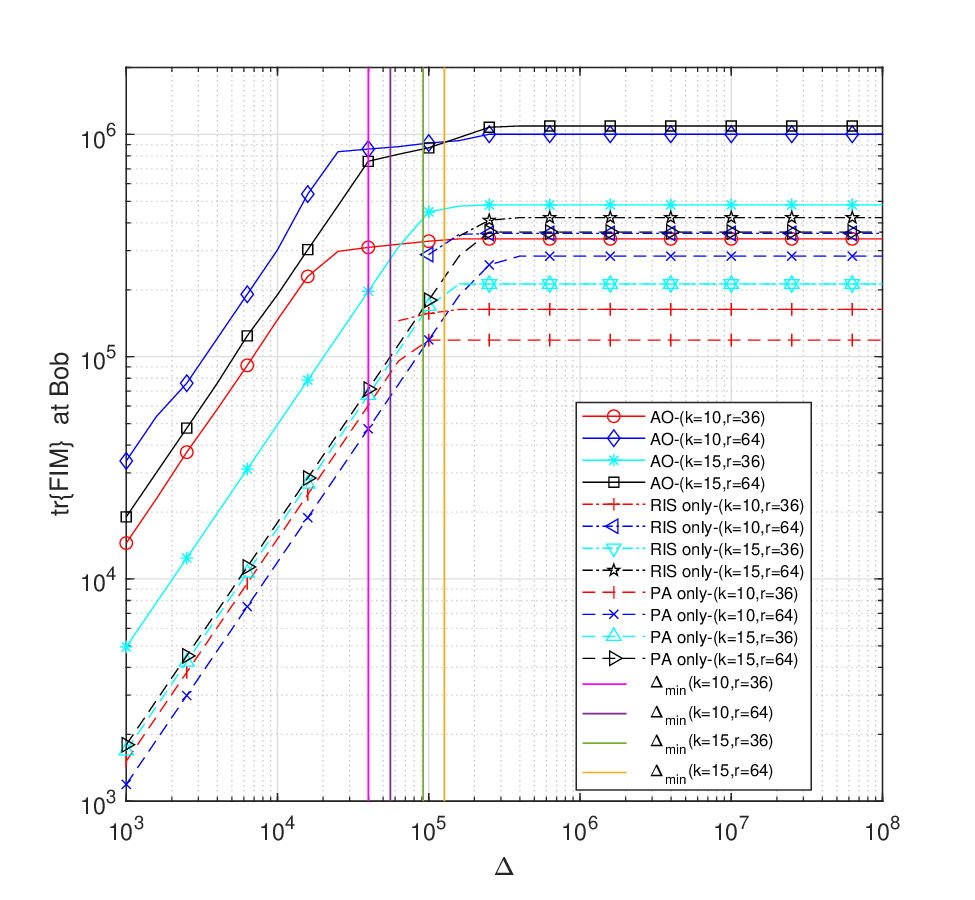}}

\vspace{-0.5cm}

\subfloat{%
  \includegraphics[width=0.9\columnwidth,draft=false]{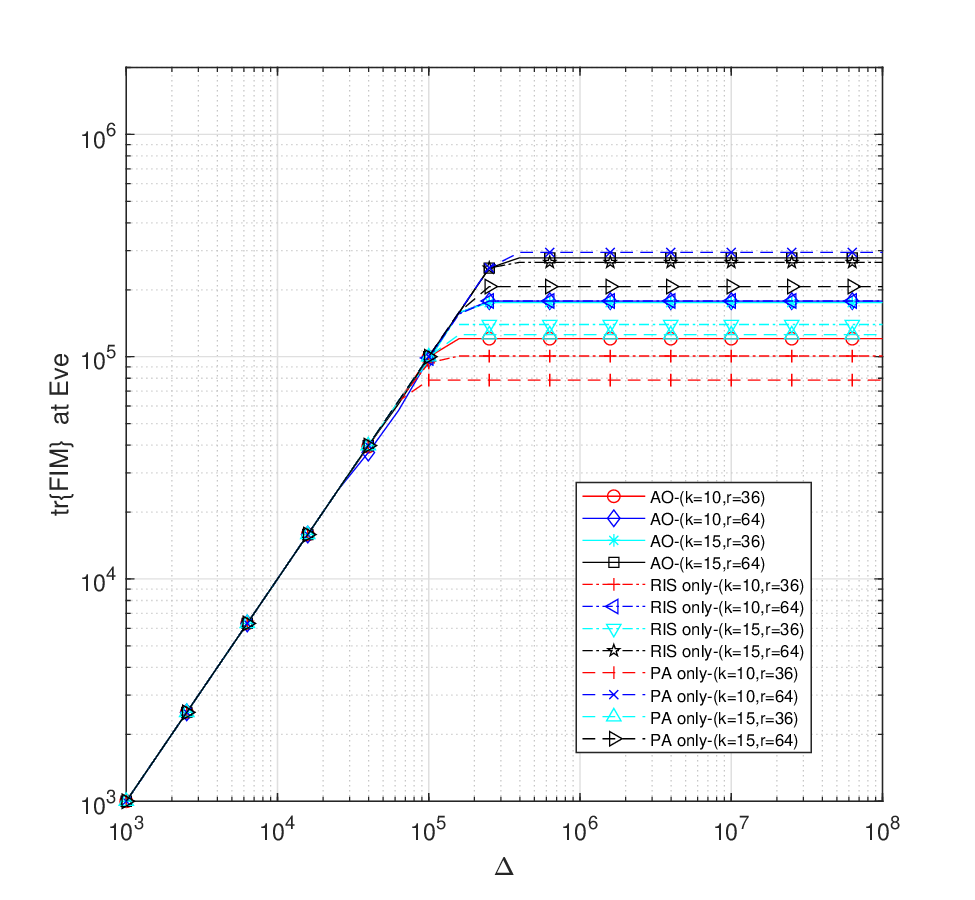}
}

\vspace{-0.1cm}

\caption{Trace of FIM achieved by the AO, RIS only, and PA only algorithms versus $\Delta$ for various values of $k$ and $r$, where $P_\Sigma=30$. Also, the feasibility analysis is included with vertical lines for each $k$ and $r$ pair.}\label{Fig:vsdelta}

\vspace{-0.35cm}

\end{figure}

First, the trace of the FIM is evaluated at Bob and Eve versus the secrecy parameter $\Delta$ for $k \in \{10,15\}$, $r\in\{36,64\}$, and $P_{\Sigma}=30$ in Fig.~\ref{Fig:vsdelta}. As expected, higher average Fisher information can be achieved at Bob when the RIS size increases. Also, it is noted that received signals at Bob are more informative than those at Eve when the secrecy constraint is satisfied. For lower values of $\Delta$, the secrecy constraint is active and the trace of the FIM is equal to the value of $\Delta$ at Eve and higher at Bob. As $\Delta$ increases, the secrecy constraint gets less restrictive and holds with inequality, which leads to certain fixed values for all of the cases depending on the system parameters. Meanwhile, the proposed AO algorithm has better performance than the benchmark algorithms, RIS only and PA only. For instance, in the case of $k=10$ and $r=64$, the information at Bob is approximately five times of that at Eve for the AO algorithm, whereas the information ratio between Bob and Eve achieved by the RIS only and PA only algorithms are approximately two and one, respectively. In addition, the RIS only algorithm has improved performance compared to PA only, which is intuitive since it has more degrees of freedom. However, this claim does not necessarily hold at Eve since the values at Eve are either equal to the secrecy constraint, $\Delta$, or depend on the structure of the system (the maximum value of ${\boldsymbol{A}}_{b}$, ${P}_{\Sigma}$, and $k$ are the parameters that determine the trace of the FIM in this scenario). Moreover, as mentioned in \eqref{eq:FeasRISoptProblem}, a feasibility region analysis takes place for the RIS only approach. In this scenario, $\Delta_{\min}$'s are derived for all the setups that are illustrated with the vertical lines in Fig.~\ref{Fig:vsdelta}. It is observed that the RIS only algorithm yields solutions when $\Delta$ is higher than these specified values. 

\begin{figure}%[htp]
\vspace{-0.4cm}

\centering

\subfloat{%
  \includegraphics[width=0.9\columnwidth,draft=false]{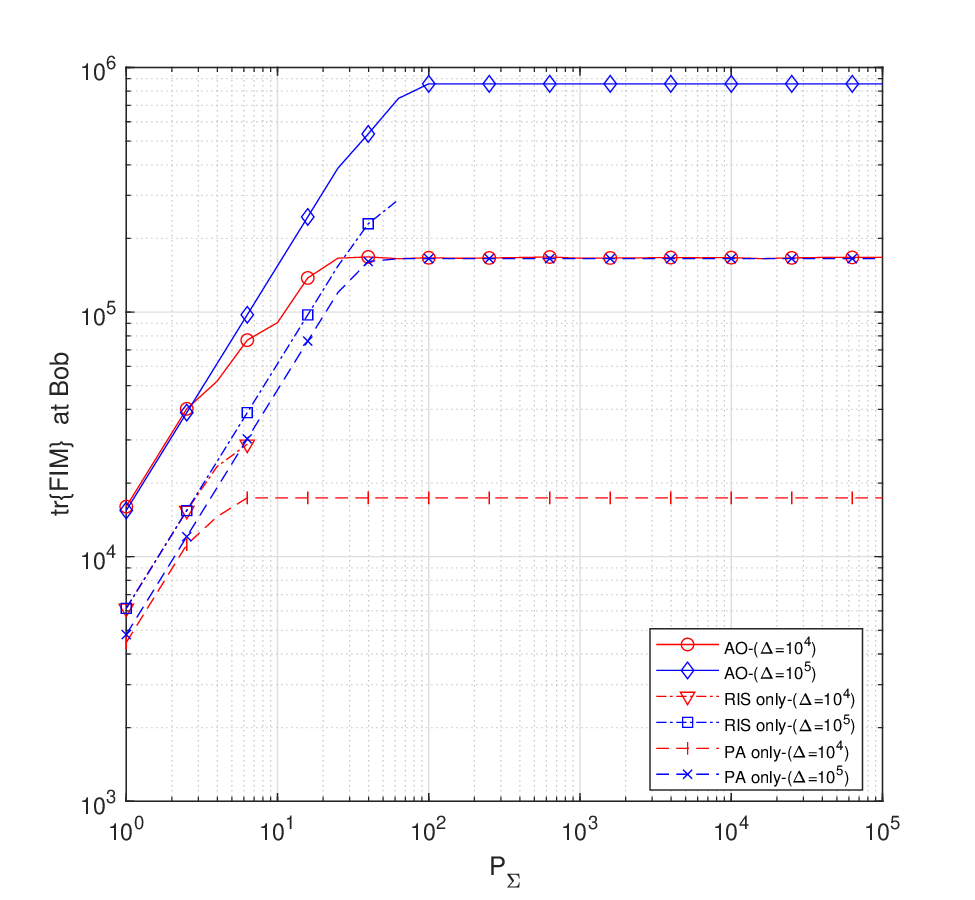}
}

\vspace{-0.45cm}

\subfloat{%
  \includegraphics[width=0.9\columnwidth,draft=false]{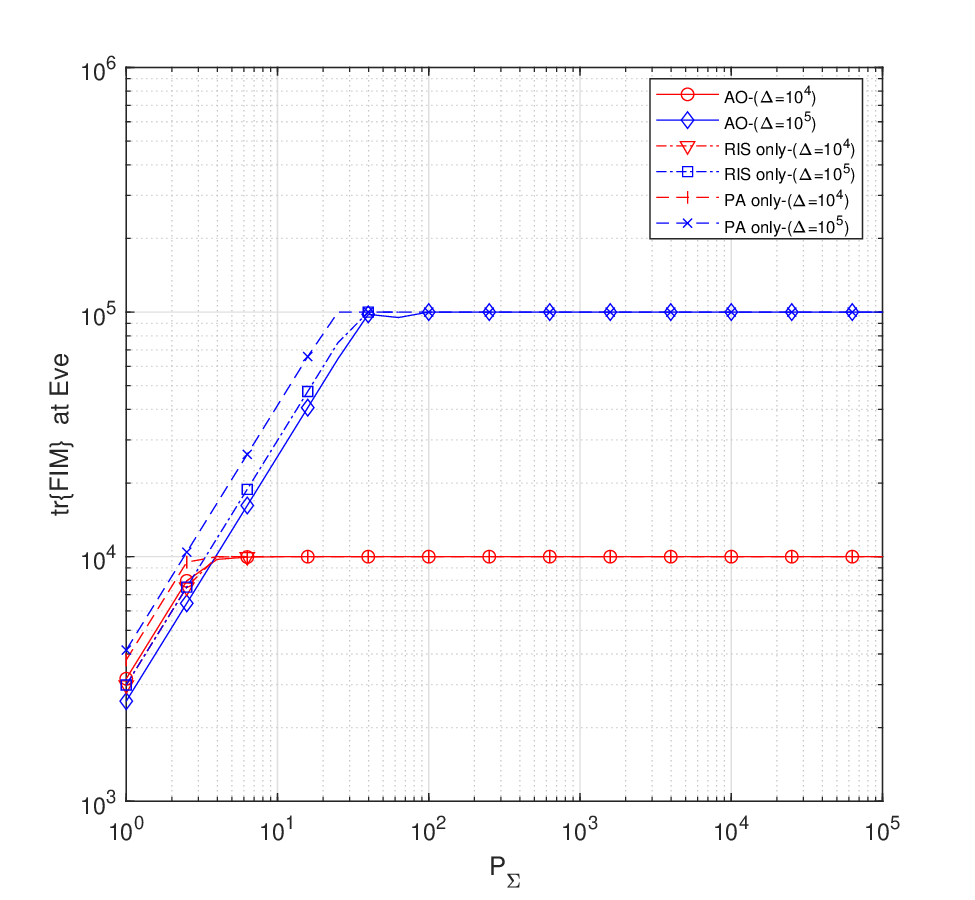}
}

\vspace{-0.1cm}

\caption{Trace of FIM achieved by AO, RIS only, and PA only algorithms versus $P_\Sigma$ for various values of $\Delta$ where $k=10$ and $r=36$.}\label{Fig:vsPsig_Delta}

\vspace{-0.35cm}

\end{figure}

\begin{figure}%[htp]
\vspace{-0.4cm}

\centering

\subfloat{%
  \includegraphics[width=0.9\columnwidth,draft=false]{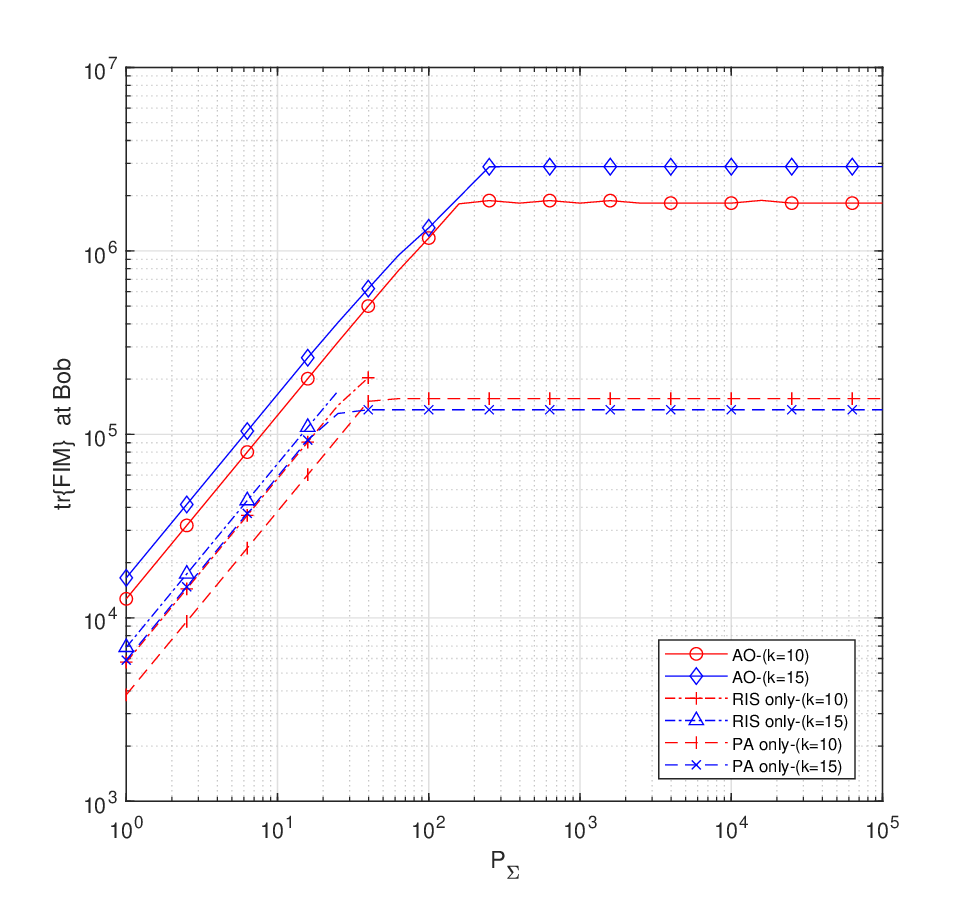}
}

\vspace{-0.45cm}

\subfloat{%
  \includegraphics[width=0.9\columnwidth,draft=false]{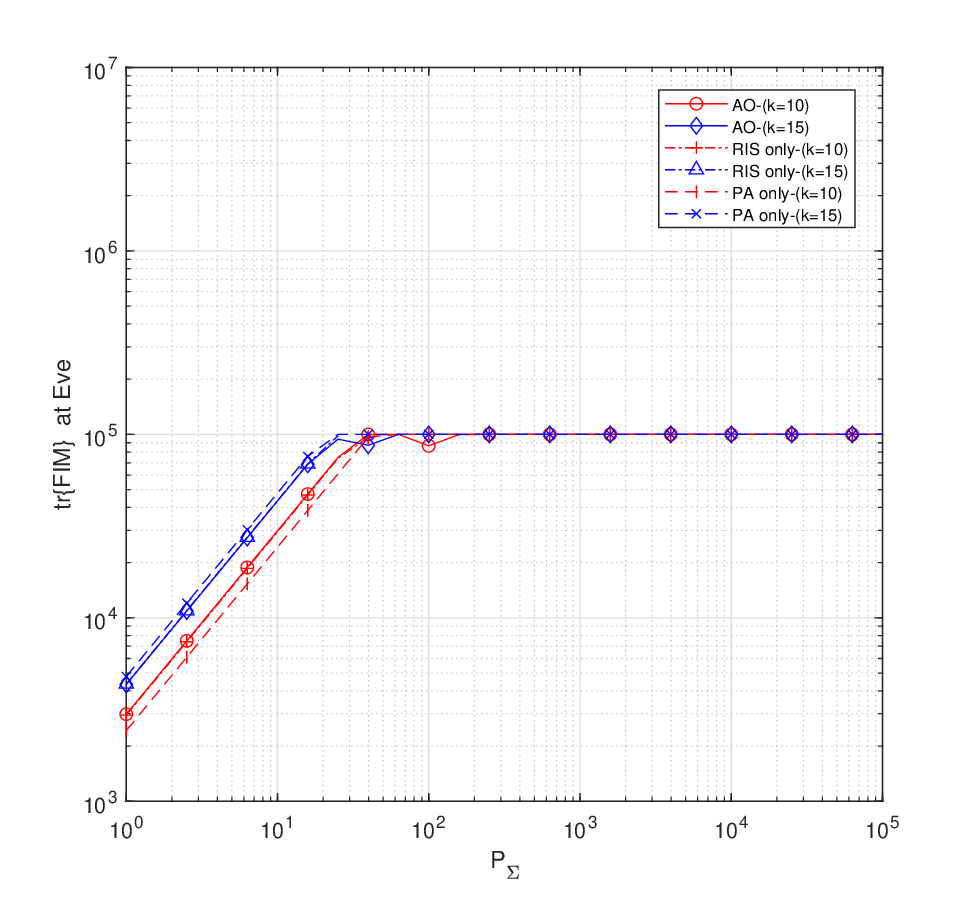}
}

\vspace{-0.1cm}

\caption{Trace of FIM achieved by AO, RIS only, and PA only algorithms versus $P_\Sigma$ for various values of $k$, where $\Delta= 10^5$ and $r=36$.}\label{Fig:vsPsig_k}

\vspace{-0.35cm}

\end{figure}

In Fig.~\ref{Fig:vsPsig_Delta} and Fig.~\ref{Fig:vsPsig_k}, the traces of the FIMs at Bob and Eve are plotted versus the total power constraint, $P_{\Sigma}$, and the effects of varying $\Delta$ and $k$ are analyzed for a fixed value of the RIS size. From Fig.~\ref{Fig:vsPsig_Delta}, where $k=10$, $r=36$, and $\Delta\in \{10^4,10^5\}$, it can be realized that as $P_\Sigma$ increases starting from low values, the average Fisher information at Bob and Eve increases up to certain levels. While the secrecy constraint is inactive for low values of $P_\Sigma$, it gets active when $P_\Sigma$ increases up to certain values. In that case, the trace of the FIM at Eve becomes equal to $\Delta$. (The RIS only approach becomes infeasible after some values of $P_\Sigma$). For both values of $\Delta$ in Fig.~\ref{Fig:vsPsig_Delta}, the AO algorithm achieves higher average Fisher information at Bob than both of the benchmarks while satisfying the secrecy constraint for Eve. Also, Bob receives more informative signals than Eve, as observed in the previous scenario. From Fig.~\ref{Fig:vsPsig_k}, where $k \in \{10,15\}$, $r=36$, and $\Delta=10^5$, it is observed that by increasing the parameter size at Alice (from $k=10$ to $k=15$), and accordingly the observation sizes at Bob and Eve (since $n_b=n_e=2k$), higher Fisher information can be achieved at Bob. 
%a certain level of performance is achieved with reduced power. %Also, the average information at Bob is larger for $k=15$ than
%that for $k = 10$ which is in compliance with the results in %Fig.~\ref{Fig:vsK}. 
In addition, similar to Fig. \ref{Fig:vsdelta}, the AO approach outperforms the benchmark algorithms, and the RIS only algorithm is more effective than the PA only algorithm in its feasible regions.
%Additionally, for smaller RIS-size, the problem is feasible for larger values of power. Hence, higher performance could be achieved if a high value of power is available for $r=50$ ($k=10$).

\begin{figure}%[htp]
\vspace{-0.4cm}

\centering

\subfloat{%
  \includegraphics[width=0.9\columnwidth,draft=false]{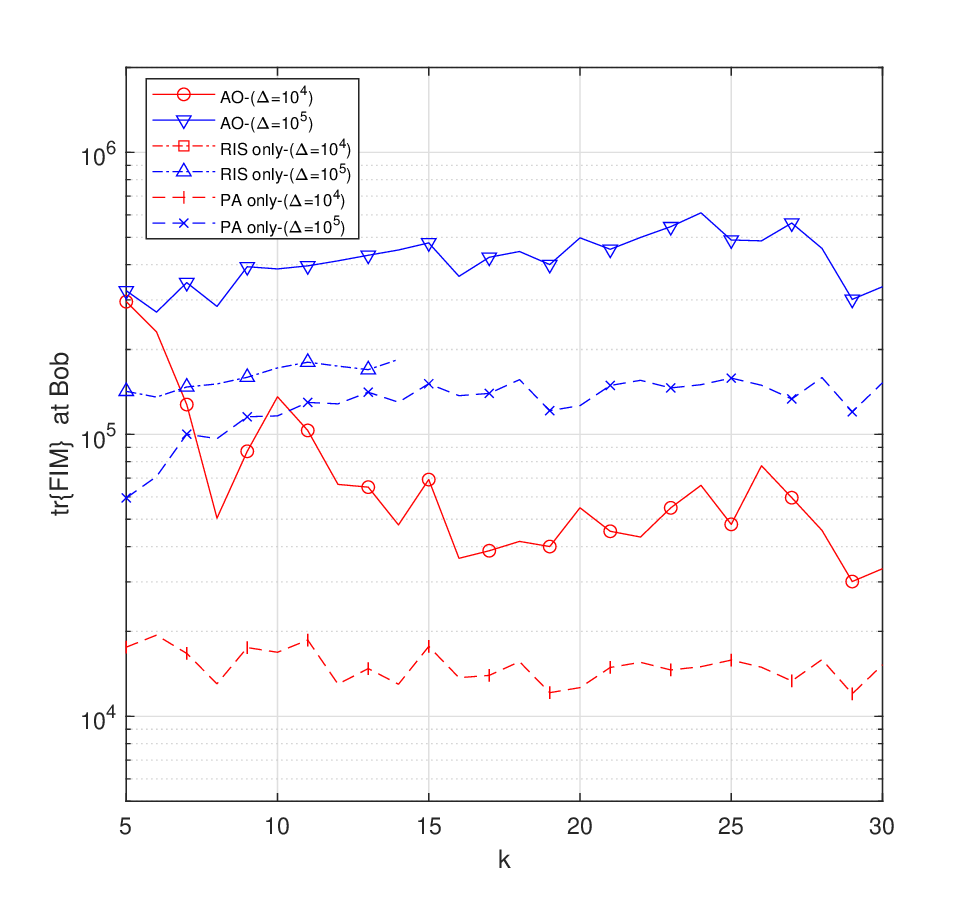}}

\vspace{-0.5cm}

\subfloat{%
  \includegraphics[width=0.9\columnwidth,draft=false]{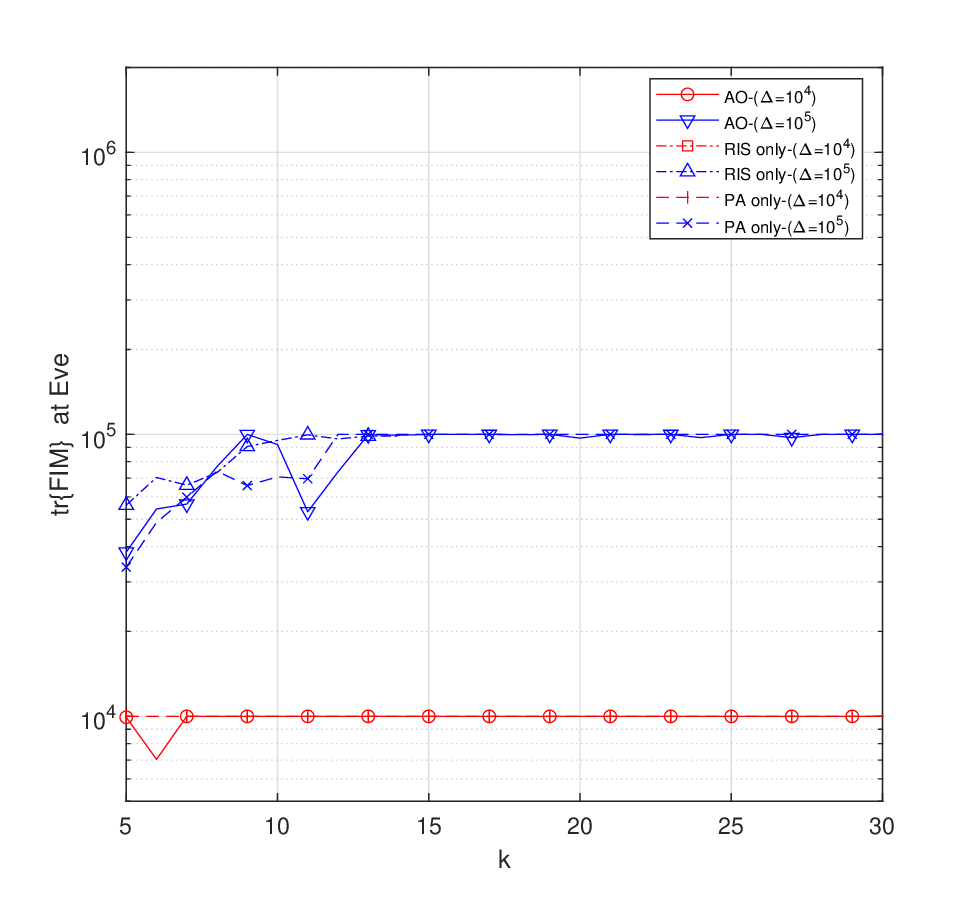}
}

\vspace{-0.1cm}

\caption{Trace of FIM achieved by AO, RIS only, and PA only algorithms versus $k$ for various values of $\Delta$, where $P_\Sigma=30$ and $r=36$.}\label{Fig:vsK}

\vspace{-0.35cm}

\end{figure}

Next, the traces of the FIMs at Bob and Eve achieved by the three aforementioned approaches are plotted versus the number of parameters, $k$, in Fig. \ref{Fig:vsK}, considering various values of $\Delta$ with $P_\Sigma=30$ and $r=36$. It is observed that as the value of the secrecy threshold, $\Delta$, increases, the received signal at Bob becomes more informative as the secrecy constraint gets less restrictive. Moreover, the average Fisher information at Bob is higher than that at Eve for every value of $k$ while maintaining compliance with the secrecy constraints. Also, at Eve, the effectiveness of the secrecy constraint is observed at all points in the figure, except for $k\in \{1,2,\dots,13\}$ with $\Delta=10^5$. Furthermore, due to the random generation of channel matrices and the direct dependence of matrices ${\boldsymbol{\mathcal{Q}}}_{b}$, ${\boldsymbol{\mathcal{Q}}}_{e}$, ${\boldsymbol{A}}_{b}$, and ${\boldsymbol{A}}_{e}$ on them, increasing the size of the parameter does not necessarily lead to an increase in the information at Bob for a given value of the total power constraint, as observed from Fig.~\ref{Fig:vsK}.
 
\begin{figure}%[htp]
\vspace{-0.4cm}

\centering

\subfloat{%

    \includegraphics[width=0.9\columnwidth,draft=false]{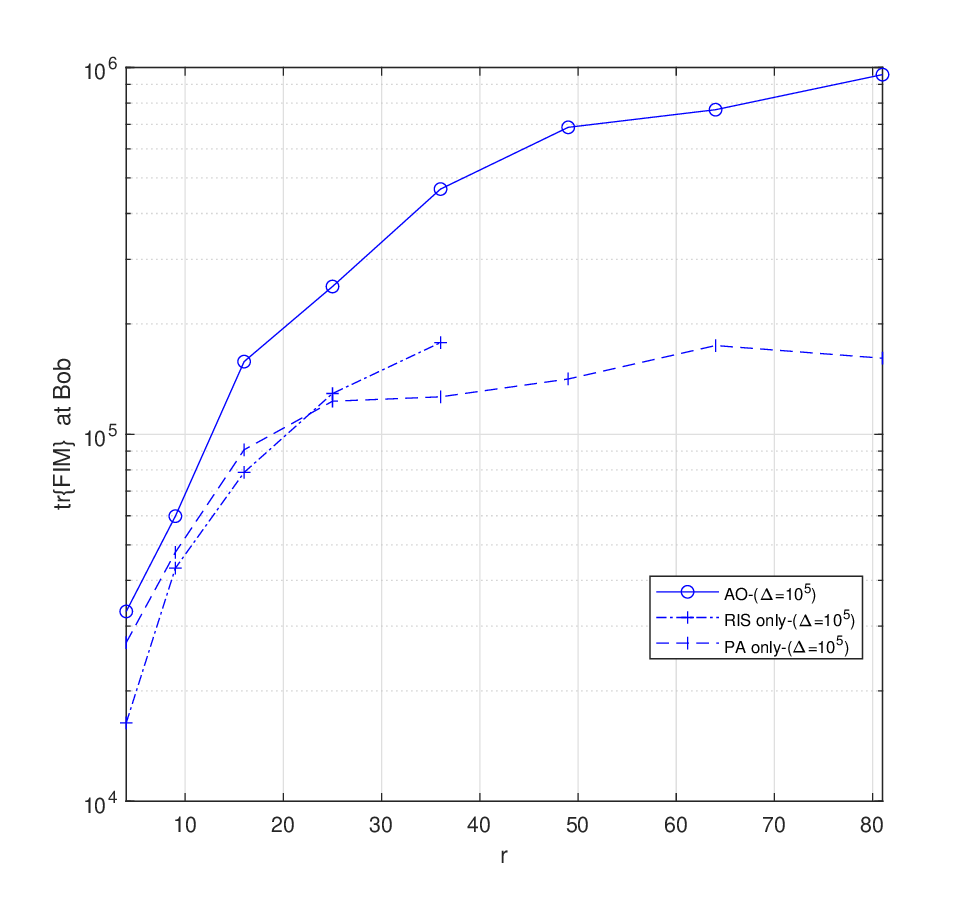}
}

\vspace{-0.5cm}

\subfloat{%
  \includegraphics[width=0.9\columnwidth,draft=false]{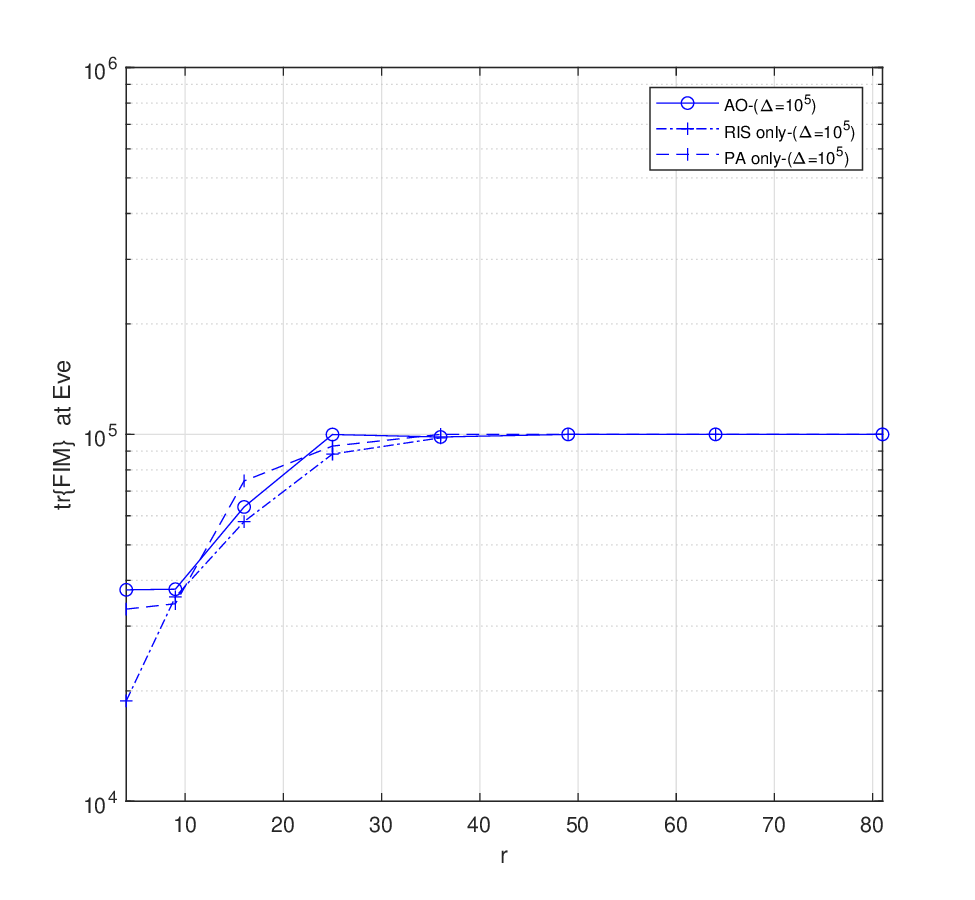}
}

\vspace{-0.1cm}

\caption{Trace of FIM achieved by AO, RIS only, and PA only algorithms versus $r$, where $P_\Sigma=30$, $\Delta=10^5$, and $k=15$.}\label{Fig:vsR}

\vspace{-0.35cm}

\end{figure}

Fig.~\ref{Fig:vsR} presents the trace of the FIM with respect to the number of RIS elements, $r$, for comparing the AO approach with the benchmarks, where $\Delta=10^5$, $k=15$, and $P_{\Sigma}=30$. The observed trends show that an increase in the value of $r$ leads to a corresponding increase in the amount of information at Bob, and the proposed AO algorithm improves the performance at Bob significantly compared to the RIS only and PA only approaches. Additionally, the RIS only approach has feasibility only for values of $r$ below $36$.
%, while the AO approach extends the capability to receive information for larger values of $r$. 
Notably, due to the presence of Eve, for the values of $r$ higher than $36$, the performance is affected by the secrecy constraint for preventing Eve from obtaining information exceeding the constraint $\Delta$.

\begin{figure}%[htp]
\vspace{-0.4cm}

\centering

\subfloat{%
   \includegraphics[width=0.9\columnwidth,draft=false]{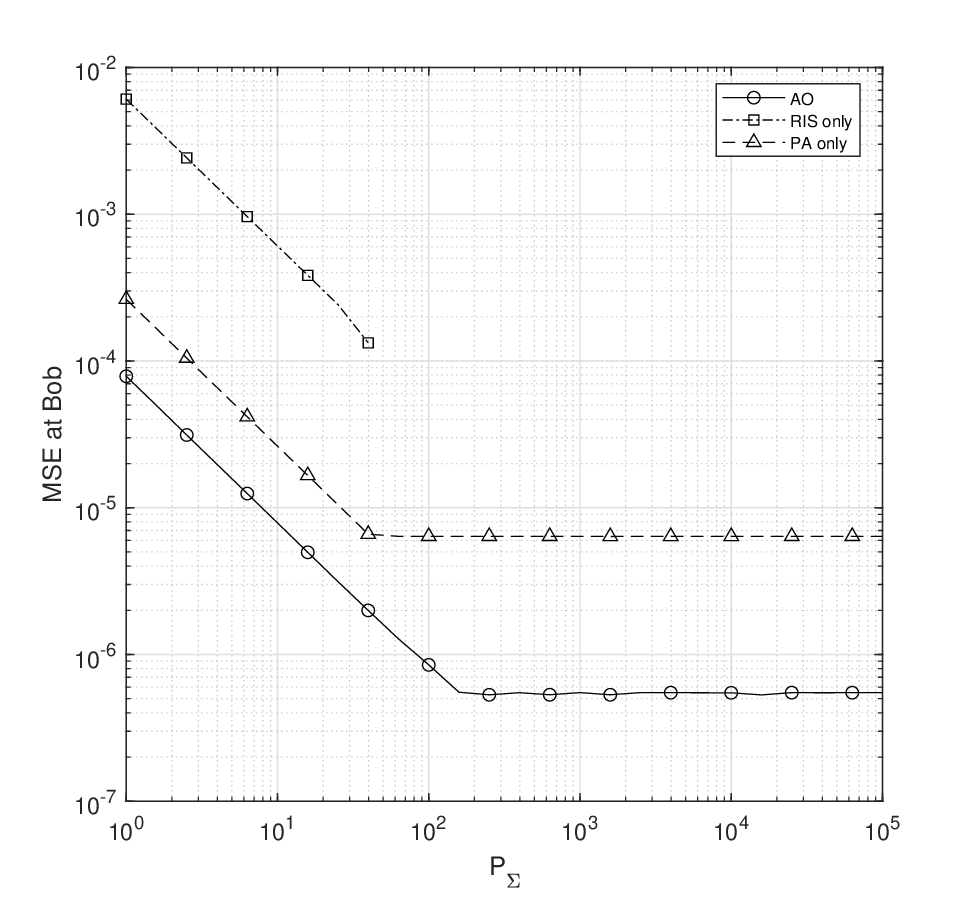}
}

\vspace{-0.45cm}

\subfloat{%
  \includegraphics[width=0.9\columnwidth,draft=false]{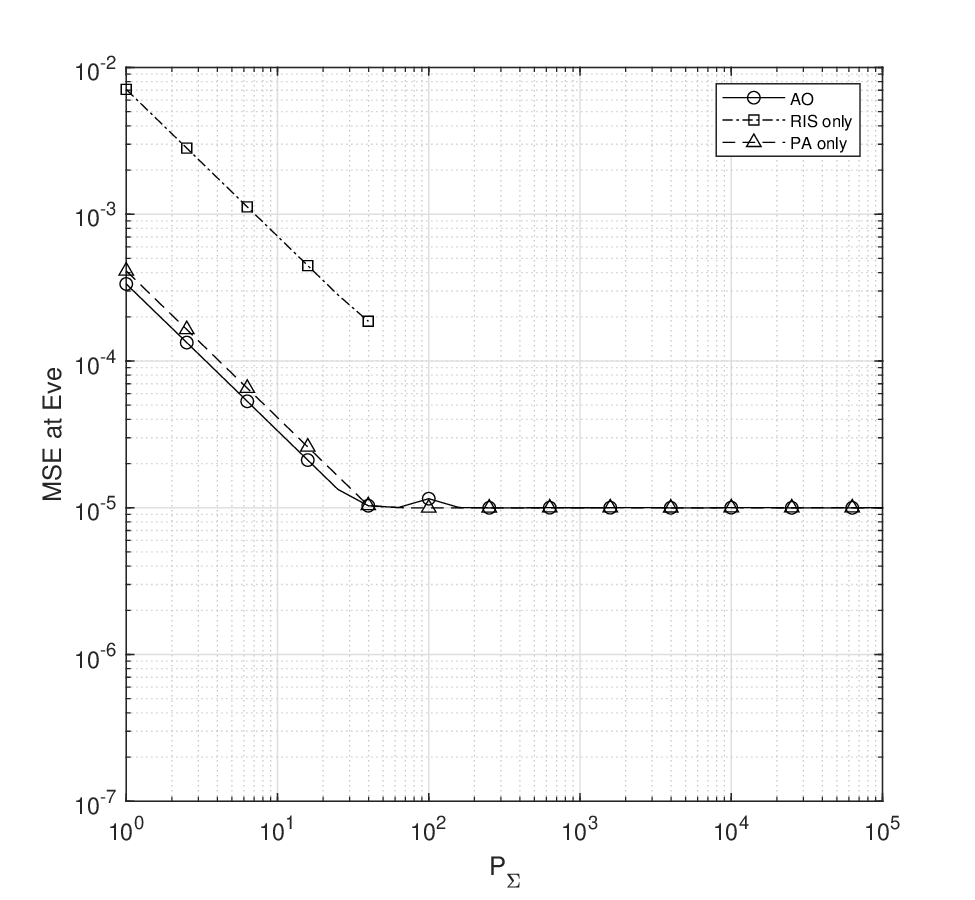}
}

\vspace{-0.1cm}

\caption{MSEs of ML estimators corresponding to the AO, RIS only, and PA only algorithms versus $P_\Sigma$, where $r=36$, $\Delta=10^5$, and $k=10$.}\label{Fig:MSE}

\vspace{-0.35cm}

\end{figure}

Finally, in order to observe the usefulness of the proposed AO algorithm for practical estimators, the MSEs of the ML estimators (MLEs) are plotted in Fig.~\ref{Fig:MSE} with respect to $P_\Sigma$ for the AO, RIS only and PA only algorithms, where $r=36$, $k=10$, and $\Delta=10^5$. In particular, the solution of the AO algorithm, as obtained from Algorithm~1, is used to set the RIS phase profile and power allocation matrices, and the resulting MSE of the ML estimator is plotted in the figure for the AO algorithm. Similarly, the RIS only and PA only solutions are obtained according to the trace of the FIM metric proposed in this study and the resulting matrices are employed when implementing the ML estimators. In calculating the MSE, $1000$ Monte-Carlo trials are performed for each setting, and the average MSE is calculated by dividing the total MSE by the number of parameters sent by Alice. (It is important to mention that in the AO and the PA only algorithms, powers of some parameters may be set to zero, meaning that they are not sent from Alice; hence, such parameters are not included in the average MSE calculation.) 
%In the PA problem, the power allocated to some parameters gets to be zero as the solution of the problem, making some rows and columns of the FIM  and, consequently, its determinant to be zero. Therefore, in order to make the FIM invertible for calculating the CRLB, we have deleted the rows and columns related to the zero-power parameters and calculated the trace of FIM$^{-1}$ which is basically the estimation accuracy of the parameters that have been transmitted via non-zero power only. 
%Similarly, we have normalized the MSE of MLE per parameter for the transmitted parameters only. Since we use a linear Gaussian structure, it can be shown that these two derivations will have similar results which could be seen in our simulation results either.
It is noted from Fig. \ref{Fig:MSE} that the proposed AO algorithm is also useful for the practical ML estimator as it leads to lower MSEs at Bob. 
%Also, for the MSE at Eve, it is shown that the secrecy constraint is satisfied as it was expected.
 
\section{Extensions to Other Scenarios}\label{sec:extensions}

\subsection{Presence of Dominant LoS Components}

Since RISs can provide significant benefits in the absence of LoS, a system model with LoS blockage is considered in the previous sections, as shown in Fig.~\ref{RIS-assisted}. In order to investigate the effects of RIS in the presence of LoS components, in this section we assume the presence of dominant LoS paths for both Bob and Eve. Accordingly, the received signals can be written, by adapting \eqref{y_bob} and \eqref{y_eve} to this scenario, as follows:
\begin{align} \label{y_bob_los}
    {\bf{y}}_b&=\left( {\bf{H}}_{rb}{\bf{\Omega}}{\bf{H}}_{ar}+{\bf{H}}_{ab} \right){\bf{P}}{\bthe}+\Eta_b\\ \label{y_eve_los}
    {\bf{y}}_e&=\left( {\bf{H}}_{re}{\bf{\Omega}}{\bf{H}}_{ar}+{\bf{H}}_{ae} \right){\bf{P}}{\bthe}+\Eta_e
\end{align}
where ${\bf{H}}_{ab}\in \mathbb{C}^{n_b\times k}$ and ${\bf{H}}_{ae}\in \mathbb{C}^{n_e\times k}$ represent the channel coefficient matrices for the Alice-Bob and Alice-Eve channels, respectively, and the other parameters are as defined in Section~\ref{sec:sys}. 

By following the same ideas as in Lemma~1 and the equations in \eqref{FIM_derivatives}--\eqref{objectivefun}, the average Fisher information related to the measurements in \eqref{y_bob_los} and \eqref{y_eve_los} can be obtained, for $i\in\{b,e\}$, as follows:
\begin{subequations}\label{objectivefun_los}
\begin{align} \label{objectivefun_losa}
    {\rm{tr}}\{  {\bf{I}}({\bf{y}}_i;{\bthe}) \} 
    &=  {\rm{tr}}\{(({\bf{H}}_{ri}{\bf{\Omega}}{\bf{H}}_{ar}+{\bf{H}}_{ai}){\bf{P}} )^H
        \\ \notag
    & {\hspace{1cm}} \boldsymbol{\Sigma}_i^{-1}(({\bf{H}}_{ri}{\bf{\Omega}}{\bf{H}}_{ar}+{\bf{H}}_{ai}){\bf{P}} ) \} \\ \label{objectivefun_losb}
    &\hspace{-0.2cm}=  {\rm{tr}} \{({\bf{H}}_{ri}{\bf{\Omega}}{\bf{H}}_{ar}{\bf{P}} )^H
        {\boldsymbol{\Sigma}}_i^{-1}({\bf{H}}_{ri}{\bf{\Omega}}{\bf{H}}_{ar}{\bf{P}} ) \} \\\label{objectivefun_losc}
    &+   {\rm{tr}}  \{{\bf{P}}{\bf{H}}_{ai}^H 
    {\boldsymbol{\Sigma}}_i^{-1}{\bf{H}}_{ai} {\bf{P}}\} \\ \label{objectivefun_losd}
    &+   {\rm{tr}}  \{{\bf{P}}{\bf{H}}_{ai}^H 
    {\boldsymbol{\Sigma}}_i^{-1} {\bf{H}}_{ri}{\bf{\Omega}}{\bf{H}}_{ar}{\bf{P}}\}   
    \\\label{objectivefun_lose}
    &+   {\rm{tr}}  \{{\bf{P}}( {\bf{H}}_{ri}{\bf{\Omega}}{\bf{H}}_{ar})^H 
    {\boldsymbol{\Sigma}}_i^{-1}{\bf{H}}_{ai} {\bf{P}}\}    
\end{align}
\end{subequations}
It is noted that the first term in \eqref{objectivefun_losb} is equal to \eqref{objectivefun}. In addition, since \eqref{objectivefun_losc} is not a function of ${\bf{\Omega}}$, it can be omitted for the RIS phase design problem. Meanwhile, the terms in \eqref{objectivefun_losd} and \eqref{objectivefun_lose} can be reformulated as in the following lemma.

\textit{{\textbf{Lemma~2:}}}
 \eqref{objectivefun_losd} and \eqref{objectivefun_lose} can be expressed as
 \begin{subequations}
     \begin{align}
     {\rm{tr}} \{{\bf{P}}{\bf{H}}_{ai}^H 
    {\boldsymbol{\Sigma}}_i^{-1} {\bf{H}}_{ri}{\bf{\Omega}}{\bf{H}}_{ar}{\bf{P}}\}
    &= \label{objectivefun_losd_L}{\boldsymbol{\omega}}^{T}{\bf{\tilde{q}}}_{i}
    \\\label{objectivefun_lose_L}
      {\rm{tr}} \{{\bf{P}}( {\bf{H}}_{ri}{\bf{\Omega}}{\bf{H}}_{ar})^H 
    {\boldsymbol{\Sigma}}_i^{-1}{\bf{H}}_{ai} {\bf{P}}\} 
    &= {\boldsymbol{\omega}}^{H}{\bf{\tilde{q}}}_{i}^*
     \end{align}
 \end{subequations}
 with ${\bf{\tilde{q}}}_i\triangleq[{\rm{tr}}\{{\bf{\tilde{S}}}_{i,1}\}, \dots, {\rm{tr}}\{{\bf{\tilde{S}}}_{i,r}\}]^T$, where ${\bf{\tilde{S}}}_{i,j}\triangleq {\bf{P}}{\bf{H}}_{ai}^H  {\boldsymbol{\Sigma}}_i^{-1} {\bf{S}}_{i,j}$ and ${\bf{S}}_{i,j}$ is as defined in \eqref{S_ij} (please see the proof of Proposition~1).

{\bf{Proof:}} Similar to the proof of \textit{Proposition~1}, define ${\bf{H}}_{i} \triangleq {\bf{H}}_{ri}{\bf{\Omega}}{\bf{H}}_{ar}{\bf{P}}$, which is given by ${\bf{H}}_{i}= \sum\limits_{j=1}^{r} \omega_j {\bf{S}}_{i,j}$ as stated in \eqref{Hi} with ${\bf{S}}_{i,j}$ being defined as in \eqref{S_ij}. Then, if we define ${\bf{\tilde{S}}}_{i,j}\triangleq {\bf{P}}{\bf{H}}_{ai}^H  {\boldsymbol{\Sigma}}_i^{-1} {\bf{S}}_{i,j}$, we can rewrite \eqref{objectivefun_losd} and \eqref{objectivefun_lose} as follows:
\begin{subequations}
\begin{align}
    {\rm{tr}}\{{\bf{P}}{\bf{H}}_{ai}^H 
    {\boldsymbol{\Sigma}}_i^{-1} {\bf{H}}_{ri}{\boldsymbol{\Omega}}{\bf{H}}_{ar}{\bf{P}}\}
    & = \sum\limits_{j=1}^r \omega_j \, {\rm{tr}}\{{\bf{\tilde{S}}}_{i,j}\}\\ 
    & \triangleq {\boldsymbol{\omega}}^{T} {\bf{\tilde{q}}}_{i}
%\end{align}
%\end{subequations}
%\begin{subequations}
%\begin{align}
    \\
    {\rm{tr}} \{{\bf{P}}( {\bf{H}}_{ri}{\bf{\Omega}}{\bf{H}}_{ar})^H 
        {\boldsymbol{\Sigma}}_i^{-1}{\bf{H}}_{ai} {\bf{P}}\} 
        & = \sum\limits_{j=1}^r \omega_j^* \, {\rm{tr}}\{{\bf{\tilde{S}}}_{i,j}^H\} \\
        & \triangleq {\boldsymbol{\omega}}^{H}{\bf{\tilde{q}}}_{i}^*
    \end{align}
\end{subequations}
where ${\bf{\tilde{q}}}_i\triangleq[{\rm{tr}}\{{\bf{\tilde{S}}}_{i,1}\}, \dots, {\rm{tr}}\{{\bf{\tilde{S}}}_{i,r}\}]^T$. Hence, the proof is completed.\hfill$\blacksquare$

Based on the results in \textit{Lemma~2} and the expression in \eqref{objectivefun_los}, the RIS phase design problem in \eqref{eq:RISomegaProblem} can be redefined in this scenario as follows:
\begin{subequations}\label{eq:NEWRISoptProblem2}
\begin{align} \label{eq:NEWRISoptProblem2a}
     \max_{{\boldsymbol{\omega}}} &\; \; \;    {\boldsymbol{\omega}}^H{\boldsymbol{\mathcal{Q}}}_{b}{\boldsymbol{\omega}}+ {\boldsymbol{\omega}}^{T}{\bf{\tilde{q}}}_{b}+ {\boldsymbol{\omega}}^{H}{\bf{\tilde{q}}}_{b}^* \\\label{eq:NEWRISoptProblem2b}
      s.t. &\; \; \;  {\boldsymbol{\omega}}^H{\boldsymbol{\mathcal{Q}}}_{e}{\boldsymbol{\omega}}+ {\boldsymbol{\omega}}^{T}{\bf{\tilde{q}}}_{e}+ {\boldsymbol{\omega}}^{H}{\bf{\tilde{q}}}_{e}^*    \leq \Delta \\ \label{eq:NEWRISoptProblem2c}
         &\; \; \,|\omega_i| = 1  \;\;\; {\rm{for}} \;\;\; i=1,2,\dots,r 
\end{align}
\end{subequations}
This inhomogeneous QCQP problem can be homogenized by adopting the approach used in \cite{SDR_QCQP} as follows:
\begin{subequations}\label{eq:inhomRISoptProblem}
\begin{align} \label{eq:inhomRISoptProblema}
     \max_{{\boldsymbol{\omega}},t} &\; \; \;
     \begin{bmatrix}
    {\boldsymbol{\omega}}^H & t
     \end{bmatrix}
        \begin{bmatrix}
    {\boldsymbol{\mathcal{Q}}}_{b} & {{\boldsymbol{\tilde{q}}}}_{b}^*\\ \\
    {{\boldsymbol{\tilde{q}}}}_{b}^T& {\bf{0}}
     \end{bmatrix}
     \begin{bmatrix}
    {\boldsymbol{\omega}} \\ t
     \end{bmatrix}\\
      s.t. &\; \; \; \label{eq:inhomRISoptProblemb}
      \begin{bmatrix}
    {\boldsymbol{\omega}}^H & t
     \end{bmatrix}
        \begin{bmatrix}
    {\boldsymbol{\mathcal{Q}}}_{e} & {{\boldsymbol{\tilde{q}}}}_{e}^*\\ \\
    {{\boldsymbol{\tilde{q}}}}_{e}^T& {\bf{0}}
     \end{bmatrix}
    \begin{bmatrix}
    {\boldsymbol{\omega}} \\ t
     \end{bmatrix}  \leq \Delta   \\ \notag
     &\; \; \; \, t^2 = 1 \\ \notag
    &\; \; \; \, |\omega_i| = 1  \;\;\; {\rm{for}} \;\;\; i=1,2,\dots,r \\   \notag  
 \end{align}
\end{subequations}
By using SDR relaxation, \eqref{eq:inhomRISoptProblem} can be solved similarly to the main RIS phase design problem in \eqref{eq:SDRRISoptProblem}.

On the other hand, for a given RIS phase profile, the optimal power allocation matrix can be found in closed-form as in Section~\ref{sec:PA_only} since the expression in \eqref{objectivefun_los} can be stated in the form of ${\rm{tr}}\{{\bf{P}}^H{\tilde{\boldsymbol{A}}}_{i}{\bf{P}}\}$ as in \eqref{eq:NEWPAoptProblem}. Namely, ${\tilde{\boldsymbol{A}}}_{i}$ becomes $\tilde{{\boldsymbol{A}}}_i \triangleq ({\bf{H}}_{ri}{\bf{\Omega}}{\bf{H}}_{ar})^H \boldsymbol{\Sigma}_i^{-1}({\bf{H}}_{ri}{\bf{\Omega}}{\bf{H}}_{ar})+{\bf{H}}_{ai}^H 
    {\boldsymbol{\Sigma}}_i^{-1}{\bf{H}}_{ai}+{\bf{H}}_{ai}^H 
    {\boldsymbol{\Sigma}}_i^{-1} {\bf{H}}_{ri}{\bf{\Omega}}{\bf{H}}_{ar}+( {\bf{H}}_{ri}{\bf{\Omega}}{\bf{H}}_{ar})^H 
    {\boldsymbol{\Sigma}}_i^{-1}{\bf{H}}_{ai}$ in this scenario based on \eqref{objectivefun_los}.

Overall, the proposed alternating optimization approach in Algorithm~1 can also be employed for the LoS scenarios as the structure of the optimization problems remains the same.

\subsection{RIS Elements with Adjustable Magnitudes}

In most studies in the literature, passive RIS elements are considered, which can control only the phase of the reflected signal, with no control over the magnitude. That is, by considering the signal reflection coefficient for each RIS element as $\omega_i = \beta_ie^{j \phi_i}$, $\beta_i=1$ is assumed for all $i=1,2,\dots,r$ \cite{RISprofile1,RISprofile2,RIS_SR2, SOP_RIS1}. According to \cite{RIS_magnitude}, via some techniques such as attaching liquid crystals or graphene under patch elements, one can change the material permeability and permittivity. However, this approach is relatively costly compared to the conventional passive RIS design. In this section, we consider the RIS phase design and power allocation problem when any of these approaches are applied to RIS and the magnitude of the reflected signal at each RIS element can be controlled independently. Since the RIS is still assumed to be passive, the signal magnitudes cannot be increased; hence, the signal reflection coefficient can be controlled to be between $0$ and $1$, i.e., $0\leq\beta_i\leq 1$. Thus, the RIS phase profile optimization problem in \eqref{eq:RISomegaProblem} can be reformulated as
\begin{subequations}\label{eq:RIS_MAG_omegaProblem}
\begin{align}\label{eq:RIS_MAG_omegaProblema}
    \max_{{\bf{\Omega}}\in\mtDr}\quad &{\rm{tr}}\{{\bf{I}}({\bf{y}}_b;{\bthe})\}\\\label{eq:RIS_MAG_omegaProblemb}
    {\rm{s.t.}}\quad\,\,\,\, & {\rm{tr}}\{{\bf{I}}({\bf{y}}_e;{\bthe})\} \leq \Delta  \\\label{eq:RIS_MAG_omegaProblemc}
    \quad & \big{|}[{\bf{\Omega}}]_{\ell,\ell}\big{|}\leq1, ~~\ell=1,\ldots,r
\end{align}
\end{subequations}
where an equality is employed instead of equality in the last constraint. 
\eqref{eq:RIS_MAG_omegaProblem} is a convex optimization problem and it does not have a feasibility issue anymore. Hence, it can be solved via CVX for any value of $\Delta$. Therefore, the alternating optimization approach proposed in this study can be applied more effectively in this scenario.

\section{Concluding Remarks}\label{sec:conclusion}

We have investigated the secure transmission of a deterministic complex-valued parameter vector from a transmitter to an intended receiver in the presence of an eavesdropper. It has been assumed that the transmitter is in NLoS with both the intended receiver and the eavesdropper, which receive the transmitted signal via an RIS available in the environment. The aim has been to design the RIS phase profile and the power allocation matrix at the transmitter to enhance the estimation accuracy at the intended receiver while limiting that at the eavesdropper. The trace of the FIM, equivalently, the average Fisher information, has been employed as the estimation performance metric, and its closed form expression has been derived. Based on that expression, the joint RIS phase profile and power allocation problem has been formulated, and it has been solved via alternating optimization. During alternating optimization, when the power allocation matrix is fixed, the optimal RIS phase profile design problem has been formulated as a non-convex QCQP, and it has been solved via SDR and rank reduction. When the RIS phase profile is fixed, an LP formulation has been obtained for optimal power allocation. Via simulations, the effects of RIS phase design and power allocation have been illustrated individually and jointly. Moreover, extensions have been presented by considering the presence of LoS paths in the environment and the availability of RIS elements with adjustable magnitudes.

%\newpage 
\bibliographystyle{ieeetr}
\bibliography{references}

\end{document}